\documentclass[journal,10pt, a4paper,twocolumns,final]{IEEEtran}
\usepackage[utf8]{inputenc}
\usepackage{amsmath,amssymb,fixmath}
\usepackage{xcolor}
\usepackage{amsthm}
\usepackage{mathtools}
\usepackage{siunitx}
\usepackage{cuted}
\usepackage{graphicx}
\usepackage{subcaption}
\usepackage{epstopdf}

\usepackage{ulem}
\usepackage{soul}

\usepackage{tabularx}
\usepackage{multirow}
\usepackage{booktabs}
\usepackage{diagbox}

\usepackage[noadjust]{cite}

\usepackage{lipsum}

\usepackage{optidef}

\renewcommand{\emph}[1]{\textit{#1}}

\RenewEnviron{BaseMiniExclam}[7]{%
	\selectConstraintMult{#1}%
	\renewcommand{\localOptimalVariable}{#2}%
	\begin{subequations}
		\ifthenelse{\equal{#7}{b}}{\allowdisplaybreaks}%
		#4
		\begin{alignat}{5}
			\bodyobj{#2}{#3}{#6}{#5}
			\BODY
		\end{alignat}
	\end{subequations}%
	\setStandardMini
}

\usepackage[norelsize,ruled]{algorithm2e}

\makeatletter
\newcommand{\removelatexerror} {\let\@latex@error\@gobble}
\makeatother

\usepackage{tcolorbox}
\tcbuselibrary{many}

\theoremstyle{plain}
\newtheorem{theorem}{Theorem}
\newtheorem{lemma}{Lemma}

\newcommand{\revise}[1]{{\color{blue}#1}}
\renewcommand{\revise}[1]{#1}

\newcommand{\superscript}[1]{^{\mathrm{#1}}}
\newcommand{\subscript}[1]{_{\mathrm{#1}}}
\newcommand{\expectation}[1]{\mathbb{E}\left\{#1\right\}}
\newcommand{\transpose}{\superscript{T}}

\newcommand{\diff}{\text{d}}
\DeclareMathOperator*{\plim}{plim}

\newtcbtheorem[]{alg}{Algorithm}{fonttitle=\bfseries}{alg}

\IEEEoverridecommandlockouts

\begin{document}

\title{Impatient Queuing for Intelligent Task Offloading in Multi-Access Edge Computing}

\author{Bin~Han, 
	Vincenzo~Sciancalepore, 
	Yihua~Xu,
	Di~Feng,
	and~Hans~D.~Schotten
	\thanks{\revise{Bin Han, Yihua Xu and Hans D. Schotten are with Technical University of Kaiserslautern, Kaiserslautern, Germany. Vincenzo Sciancalepore is with NEC Laboratories Europe, Heidelberg, Germany. Di Feng is with University of Lausanne, Lausanne, Switzerland. Bin Han (bin.han@eit.uni-kl.de) is the corresponding author.}}
}



\maketitle

\begin{abstract}
Multi-access edge computing (MEC) emerges as an essential part of the upcoming Fifth Generation (5G) and future beyond-5G mobile communication systems. {It adds computational power towards the edge of cellular networks, much closer to energy-constrained user devices, and therewith allows the users to offload tasks to the edge computing nodes for low-latency applications with very-limited battery consumption}. However, due to the high dynamics of user demand and server load, task congestion may occur at the edge nodes {resulting in long queuing delay}. Such delays can significantly degrade the quality of experience (QoE) of some latency-sensitive applications, raise the risk of service outage, and cannot be efficiently resolved by conventional queue management solutions.
	
In this article, {we study a latency-outage critical scenario, where users intend to limit} the risk of latency outage. We propose an impatience-based queuing strategy for such users to intelligently choose between MEC offloading and local computation, allowing them to rationally renege from the task queue. The proposed approach is demonstrated by numerical simulations {to be efficient for generic service model}, when a perfect queue {status} information is available. For the practical case where the users obtain {only imperfect} queue {status} information, we design an optimal online learning strategy to enable its application in Poisson service scenarios.

\end{abstract}


\IEEEpeerreviewmaketitle


\section{Introduction}\label{sec:intro}

While the maturity of current microelectronic technology enables portable devices to accomplish severe and complex tasks, there is a hype on the cloud computing services as enabler to offload such a burden onto ad-hoc cloud server---usually located at the edge of the network---based on the specific service requirements. This has recently fostered a novel paradigm shift towards the Edge Computing concept~\cite{OLC+2019MEC}.

Edge computing (or Multi-access Edge Computing, MEC) represents the first-mover advantage for the upcoming generation of network design (5G) and the core feature of beyond-5G (B5G) networks, as it brings computation resources closer to the user equipment (UE). This enables to offload computation intensive tasks to the local edge computing node instead of the centralized mobile cloud computing (MCC) server, so that the latency can be significantly reduced~\cite{MB2017MEC}. With the ever-increasing demand for computing, MEC is becoming an essential solution for expected B5G use cases with very-stringent latency requirements, where a long response time of computation task may result in degraded utility or, in the worst case, a service disruption, e.g., when dealing with ``age-of-information'' or ``age-of-task'' sensitive services~\cite{YTHR2017timely,SQT+2019age}.

Though widely recognized as an effective solution to resolve congestion in the central cloud, MEC itself may also be challenged by task congestion due to the complex load dynamics in realistic MEC scenarios. Various events, such as an unexpected burst of arriving computing tasks at the MEC server, several atypically time-consuming tasks, or a temporary reduction in computing capacity of the MEC server due to cross-slice resource scheduling, can overload the MEC server in a short term, and therewith result in an increased latency. Especially in the context of B5G/6G, with the prediction $5$-folded increase in mobile traffic over the next decade, and the 6G ambition of all-round performance enhancement w.r.t. 5G~\cite{JHHS2021road}, MEC congestion is not remaining a rare situation like believed so far, but becoming an emerging issue that cannot be overlooked.

%
%
There have been initial efforts made on this challenge in the recent past~\cite{LGLL2018deep,LYL+2019energy,HLW+2020cloud,LWW+2021joint}, which look for intelligent MEC offloading decision making mechanisms that mitigate MEC congestion, taking into account of the UE energy consumption and time delay. However, they generally consider only a static service scenario, where the latency is simplified to a consistent and universal (or average) value for all tasks. In practical mobile networks, to contrary, the latency is a random variable that is highly sensitive to the instantaneous server load. Furthermore, they mostly aim at optimizing the average power/latency performance over the network, while it is the outage-risk control making more critical impact in many future B5G applications, such as remote control and automation~\cite{HYJ+2020robustness}.

Taking another point of view, the problem of MEC server overload can also be investigated in perspective of queue congestion control. The problem of solving queue congestion has a long history in the research area of networking. A number of approaches have been proposed in the literature, such as the active queue management (AQM) technique based on queue truncation, active dropping, and preemption. AQM is proven to be the most effective approach in data networking applications such as packet routing and switching: it is capable to guarantee a full utilization of network resources, even when specified with a simple fair dropping policy~\cite{BFP09law}. Nevertheless, despite its mature development over three decades and wide success in data networking, there has been only very limited efforts such as \cite{ISS+2019congestion} to deploy AQM in cloud computing and MEC. One significant reason is the fundamental difference between the natures of packet-level and service-level queue management: while dropping and preemption of packets in data links can be easily remedied by automatic repeat request (ARQ) without compromising the quality of data traffic, denial of awaiting requests for cloud computing task or cancellation of an ongoing task will usually terminate the service session, and even reduce the user's future interest in the cloud service provider. This calls for a perspective change where a deep analysis of user behaviors may benefit the overall service queue management, as we have recently suggested in~\cite{HSC+2020multiservice}: a detailed analysis of the well-known \emph{impatient queuing} concept might help to prevent unexpected service disruptions while pushing for ad-hoc scheduling policies. Specifically, rational users may naturally behave impatiently if the waiting time exceeds reasonable levels (and thus failing to generate the expected end utility), and thereby hesitate to enter the queue or renege on its entrance into the queue and leave before being served~\cite{Stanford1979reneging}. 

To address these issues, in this paper we pioneer a risk-sensitive MEC offloading decision solution that relies on the users' impatient behavior analysis. To summarize, \emph{i)} given an arbitrary latency distribution, we have modeled the latency-sensitive utility of a computing task, and considered the user to make risk-based offloading decision between the local computing and MEC options, \emph{ii)} we have proposed a risk-based impatience mechanism that allows the users to flexibly regret its previous decision, and studied the conditions to exclude such regretting behavior, \emph{iii)} we have devised optimal strategies of regretting in two typical and realistic scenarios where the aforementioned conditions are not fulfilled and, \emph{iv)} we have validated our proposal by means of an exhaustive numerical evaluation campaign.

Compared to existing solutions, our proposal is novel regarding the following aspects: \begin{itemize}
	\item {While most existing approaches are aiming at increasing the mean utility of tasks, we propose to make task offloading decisions with respect to the risk of latency outage, i.e. the probability of \emph{remaining} waiting time exceeding a certain threshold.}
	\item We consider impatient behavior of users in queues, which is rarely studied in this field but can be important for latency-critical services.
	\item In addition to the ideal case of Poisson (Markovian) service and perfect information of the queuing system, which is commonly investigated by literature, we consider more generic cases of i) Markov-modulated Poisson service, and ii) partial system information. 
\end{itemize}

The remainder of this paper is organized as follows: we begin with a brief review to different fields related to our study in Section~\ref{sec:related}, then in Section~\ref{sec:mechanism} we set up the investigated problem, and formerly discuss about the decision model of risk-based \revise{user} impatience. We then analyze the user risk preference in more depth with Sections~\ref{sec:perfect_qsi}, proving that users will never regret in Poisson service queues, when a perfect queue status information (QSI) is available. {In the same section, subsequently, we push the discussion by one step further into the cases of non-Poisson service, where a rational user may regret from its own former decision upon updated belief in the \revise{QSI}. This impatient behavior is then proven in Section \ref{sec:poisson_queues_with_imperfect_qsi} to occur also in Poisson queues, when only an imperfect QSI is available to the users. For both cases of impatience, we derive} the optimal regretting strategies correspondingly. The proposed approaches are then numerically evaluated in Section~\ref{sec:sim} and compared against baseline solutions, before providing concluding remarks and some outlook in Section~\ref{sec:conclusion}.

\section{Related Work}\label{sec:related}

Research on the client impatience in queuing systems dates back to the 1950s, when the client behavior of  hesitating to join long queues was first noticed~\cite{Kawata1955problem}, and afterwards rigorously investigated~\cite{KR1957calculation}. The most important pioneering work in this field was conducted by \emph{Haight}, where the aforementioned phenomenon was named as ``balking''~\cite{Haight1957queueing}, and later generalized to the ``reneging'' concept~\cite{Haight1959queueing}. In 1963, \emph{Ancker} and \emph{Gafarian} pushed the work of \emph{Haight} further, proposing several most widely used statistical models of balking and reneging, and therewith analyzing their impacts on $M/M/1$ queues~\cite{AG1963some,AG1963some2}. A generalization of these analyses was presented in \cite{Stanford1979reneging} regarding $GI/G/1$ queues. 

Recently, in the context of network slicing, which is an emerging use case of public cloud computing in the Fifth Generation (5G) wireless networks, we have revisited the topic of client impatience from the micro-economic perspective in a recent work~\cite{HSC+2020multiservice}. We demonstrated how the statistic models of balking and reneging in $M/M/1$ queues are determined by the distribution of reward generated in the cloud service, and how this behavior will impact the resource utilization of multiple heterogeneous queues sharing a same server. Additionally, we also discussed the impact of clients' knowledge level in their decisions of balking and reneging.

Despite the commonness of assuming Poisson arrivals and services in literature of cloud computing, the validity of such assumption can be questionable in real service scenarios. Regarding complex real-world dynamics where the Poisson model fail to fit, Poisson \revise{mixture.} models have been intensively studied since the 1990's. Readers with specific interest in Poisson mixture models are referred to~\cite{Ryden1994parameter,CG1995poisson,JZ2005generalized}.

The problem of optimal Poisson learning, which plays an important role in risk-based reneging when only imperfect status information of the Poisson queue is available (which will be studied in Section~\ref{sec:poisson_queues_with_imperfect_qsi}), is strongly related to the research branch of decision making on the classic problem of ``when to stop'' in a learning process. Investigations to the optimal learning duration problem were initiated in the 1940s~\cite{Wald1947foundations,ABG1949bayes}, extended at the beginning of this century\cite{MS2001optimal}, and recently revisited \cite{CM2019optimal,Zhong2019optimal}.

In~\cite{TPD_Zhao21}, the authors introduce a model for offloading tasks with service caching by assuming that UEs can deploy the corresponding edge services onto the edge cloud before offloading their tasks, which are supposed to have some dependency among each other. Moreover, they consider the option of relaying the task offloading request to a further edge server if the closest one cannot meet the requested computational capacity requirements. 
The authors of~\cite{JSAC_alem19} make similar assumptions by studying a network of edge servers and allowing task offloading request relay, but they differentiate by accounting for delay requirements while disregarding any possible dependencies among tasks. Within the scope of task offloading, repetitive tasks stand out driven by the ubiquity of wireless sensors. Specifically, sensors may generate computationally intensive tasks in a periodic fashion, i.e. according to their duty cycle. \cite{TON_sl20} develops a game theoretical model and derives a polynomial time decentralized algorithm whereby UEs autonomously decide whether to offload their task or not in the current periodic slot. 
In~\cite{Access_moh20}, the authors disclose an algorithm to perform load balancing for IoT applications in the edge computing server with the aim of reducing the overall application response time. In particular, they consider the communication cost, computation time and average load on each edge computing server and develop two evolutionary algorithms, namely ant colony optimization and particle swarm optimization, to find a sub-optimal solution of the problem, which is proved to be NP-hard. The model assumes that each IoT node generates data following a Poisson distribution and employs traffic theory concepts to derive time-averaged performance figures, thereby considering full knowledge of static tasks and disregarding the task admission phase. Moreover, the authors do not envision for the use of machine learning to grant or reject task-offloading requests.

\section{Mechanism Design}\label{sec:mechanism}
\subsection{Task Offloading and Queuing Model}
In this article we consider a generic MEC task offloading scenario, where numerous users independently and randomly generate computing tasks. When a user generates a new task, it can choose between two options: either to carry it out locally on the user device, or to offload it onto the MEC server. When locally executing the task, it takes the user a stochastic latency to obtain the result, and we consider the user to have perfect knowledge about this stochastic latency. When relying on the MEC, first it always takes a certain minimal latency $\tau_0$ as overhead to accomplish the data transmission and signaling between the user and the MEC server. In addition to the that, all tasks offloaded by different users wait in a single task queue, and it takes every task an extra random waiting time in the service queue, until it is processed by the server. {Similarly, we consider a perfect knowledge about this random queuing delay to be possessed by the MEC server. Such knowledge, either for local processing delay or MEC queuing delay, must be acquired through an accurate estimation of task execution time, which has been well studied as a key enabling technology of cloud computing regarding load balancing~\cite{CSL2016load,FLHE2014load} and workflow scheduling~\cite{CBK+2017execution}. Various approaches have been proposed to address this issue on different levels~\cite{GMH2001reliable,BFSS2001source}}.

Without loss of generality, we consider the pending tasks to be processed by the server according to the ``First Come, First Served'' (FCFS) policy. The analyses and proposals in this article can be conveniently adapted to alternative queuing policies, such like ``Last Come, First Served'' (LCFS), random selection for service (RSS), and priority-based (PR), by replacing the stochastic model of waiting time, which is omitted in this {work}. For a generic mechanism design, we consider the arrivals of offloaded tasks and the service of pending tasks as independent random processes, without assuming specific distribution of inter-arrival interval or inter-service interval. We also consider the task queue to be non-preemptive with an infinite capacity. The queue offloaded tasks are usually considered in literature to be served by $c>1$ independent servers. For the convenience of analysis, in this work we simplify the case into a single server ($c=1$). Thus, in summary, we are studying a $G/G/1/\infty$ FCFS non-preemptive queue, which can be easily extended into a more general $G/G/c/\infty$ case where $c>1$.

Specially, we consider the users to be possibly \emph{impatient} in queuing. More specifically, having chosen the MEC option and waiting in the service queue, a user can regret from its previous decision at any time, retract its task from the queue without being served, and execute it locally instead. This can be identified as the well-known \emph{reneging} behavior in queuing theory~\cite{Haight1959queueing}. Similarly, the decision to locally carry out the task can also be understood as the \emph{balking} behavior where the user refuses to join the queue~\cite{Haight1957queueing}.

\subsection{Latency-Sensitive Utility Model}
Every task, after being successfully processed, will generate a utility~{(reward)} for the user that issues it. Without loss of generality, here we consider a latency-sensitive model, in which the end utility $u$ monotonically decays along with the service latency $\Delta t$, asymptotically approaching 0 while staying positive, i.e:
\begin{align}
	&u(\Delta t)\geqslant 0,\quad {\partial u}/{\partial\Delta t}<0,\quad\forall \Delta t\geqslant 0;\\
	&\lim\limits_{\Delta t\to+\infty}u(\Delta t)=0
\end{align}
where $\Delta t$ is the \emph{total} delay from the generation to the completion of the task, and $u(0)=u_0>0$ is the basic utility without latency discount.

We assume every user knows the $u_0$ of its own task a priori, so that at any given time $t$, the cumulative density function (CDF) of $\Delta t$, namely $F_{\Delta t}$, depends on the offloading decision $z_t$:
\begin{equation}
	F_{\Delta t}(x\vert t)=\begin{cases}
		F_{\tau\subscript{l}}(x-t\vert t)&z_t=0;\\
		F_{\tau\subscript{c}}(x-t\vert t)&z_t=1,
	\end{cases}
\end{equation}
where $z_t=0$ and $z_t=1$ denote local computing and MEC offloading, respectively. The instantaneous CDF $F_{\tau\subscript{l}}(x\vert t)$ of \emph{remaining} local computing delay $\tau\subscript{l}$ at $t$ is known by the user, while the instantaneous CDF $F_{\tau\subscript{c}}(x\vert t)$ of \emph{remaining} MEC computing delay $\tau\subscript{c}$ is known by the MEC server. Generally, there shall be a mechanism for the user to obtain/update the knowledge about $F_{\tau\subscript{c}}(x\vert t)$, either perfect or imperfect, from the MEC server.

\subsection{Risk-Based Impatient Queuing at MEC Server}
Interestingly, in this work we investigate a risk-critical scenario, where the maximal achievable utility w.r.t. a certain fixed outage risk $P\subscript{o}$ shall be maximized by (re-)making the offloading decision:
\begin{equation}\label{eq:offloading_decision}
	\begin{split}
		z_t=&\begin{cases}
			0&\max \left\{u(x):1-F_{\tau\subscript{l}}(x\vert t)\leqslant P\subscript{o}\right\}\\
			 &\geqslant \max\left\{u(x):1-F_{\tau\subscript{c}}(x\vert t)\leqslant P\subscript{o}\right\}\\
			1&\max \left\{u(x):1-F_{\tau\subscript{l}}(x\vert t)\leqslant P\subscript{o}\right\}\\
			 &<\max\left\{u(x):1-F_{\tau\subscript{c}}(x\vert t)\leqslant P\subscript{o}\right\}\\
		\end{cases}\\
		=&\begin{cases}
			0&\min \left\{x:F_{\tau\subscript{l}}(x\vert t)\geqslant 1-P\subscript{o}\right\}\\
			&\leqslant \min\left\{x:F_{\tau\subscript{c}}(x\vert t)\geqslant 1-P\subscript{o}\right\}\\
			1&\min \left\{x:F_{\tau\subscript{l}}(x\vert t)\geqslant 1-P\subscript{o}\right\}\\
			&>\min\left\{x:F_{\tau\subscript{c}}(x\vert t)\geqslant 1-P\subscript{o}\right\}\\
		\end{cases}\\
		=&\begin{cases}
			0&F^{-1}_{\tau\subscript{l}}(1-P\subscript{o}\vert t)\leqslant F^{-1}_{\tau\subscript{c}}(1-P\subscript{o}\vert t)\\
			1&F^{-1}_{\tau\subscript{l}}(1-P\subscript{o}\vert t)>F^{-1}_{\tau\subscript{c}}(1-P\subscript{o}\vert t)
		\end{cases},
	\end{split}
\end{equation}
where $F^{-1}_{\tau\subscript{l}}(x\vert t)$ and $F^{-1}_{\tau\subscript{c}}(x\vert t)$ are the $t$-instantaneous inverse CDFs of remaining local and edge computing latency, respectively. As the knowledge of $F_{\tau\subscript{l}}(x\vert t)$ and $F_{\tau\subscript{c}}(x\vert t)$ vary as functions of $t$, a user may repeatedly flip its choice between local and edge computing. In practice, the latency of MEC is usually more dynamic than the local computing one, because the MEC latency is not only determined by the duration of computing the task itself, but also by the queuing latency and the network delay. Therefore, here we investigate an arbitrary task with a consistent $F_{\tau\subscript{l}}(x\vert t)\overset{\forall t}{\equiv}F_{\tau\subscript{l}}(x)$ but inconsistent $F_{\tau\subscript{c}}(x\vert t)$. Without loss of generality, in the remainder part of this article, when discussing a single computing task request, we always consider it to be issued at $t=0$, if not explicitly stated otherwise. 

{Furthermore, to transfer the MEC server's statistical knowledge of queuing status $F_{\tau\subscript{c}}(x\vert t)$ to every awaiting user in queue, there must be an additional communication mechanism. In the context of MEC, to avoid waste of radio resources, it is neither practical to hold an active RRC connection for every waiting user just to update such knowledge, nor efficient to set up a new RRC connection upon every update. One reasonable solution is to exploit the paging channel for sending the QSI to the users, which does not take any extra user plane bandwidth but a paging cost. This paging cost can be restricted to an affordable level when the QSI is only sent to a small group of users, e.g. the ones awaiting in the MEC server's task queue. However, if we want to enable a symmetric reneging, i.e. a user can also regret from its decision of local computing at any time and choose to offload its task to the MEC again, our proposed mechanism will essentially demand that every user, no matter if it is waiting in the queue or processing the task locally, keeps observing the queue status at MEC server. In this case, the QSI $F_{\tau\subscript{c}}(x\vert t)$ must be \revise{broadcast} to \emph{all} users under coverage of the MEC server, which leads to a significantly increased paging cost and reduce the resource efficiency. Due to this reason, in this work} we consider an asymmetric {regretting} procedure: a user is able to {retrieve} its offloaded task from the MEC queue at any time, but never regrets from a decision of local computing. In other words:
\begin{equation}
	z_t=\revise{0}\Rightarrow z_{t'}=\revise{0},\forall t'>t.
\end{equation}

The generalization to a more generic case, where $F_{\tau\subscript{l}}(x\vert t)$ is inconsistent and the user can preempt a locally undergoing task and offload it to the MEC server, will be straightforward.

\begin{figure}
    \centering
    \includegraphics[width=\linewidth]{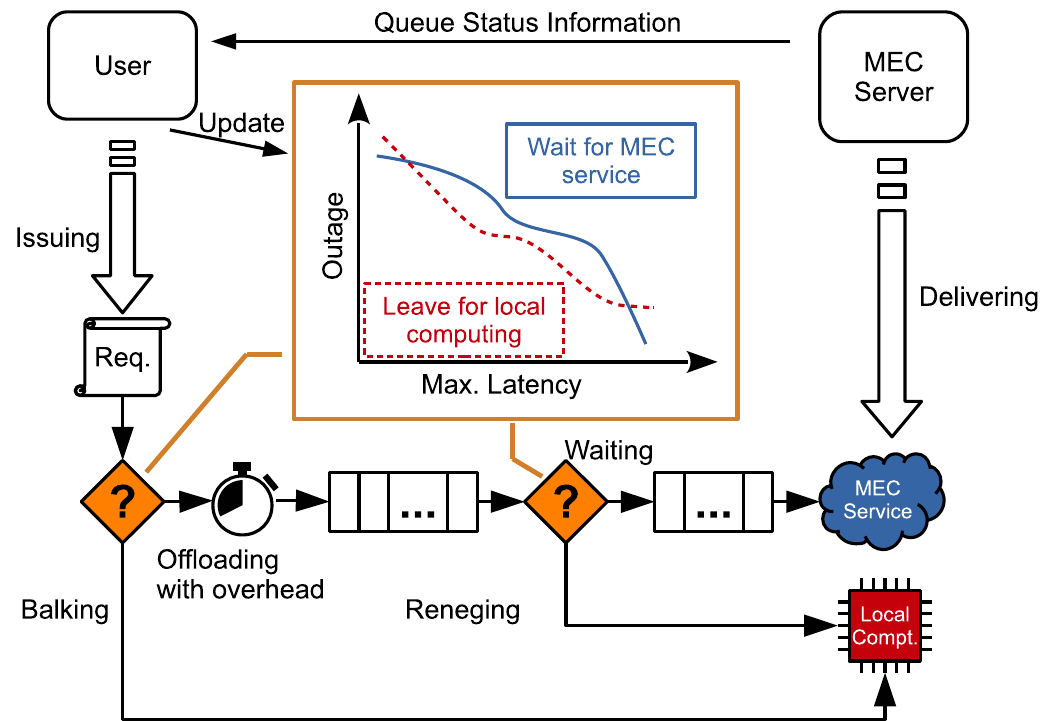}
    \caption{An overview of the proposed risk-based \revise{user} impatience mechanism}
    \label{fig:mechanism_overview}
\end{figure}

Especially, under this setup, when $z_0=\revise{0}$, the task never joins the waiting queue at MEC server, which is the balking phenomenon; when $z_0=\revise{1}$ but $z_t=\revise{0}$ for some $t>0$, the task joins the queue first and then leaves before being served on MEC server, which is the reneging phenomenon. This mechanism can be briefly illustrated by Fig.~\ref{fig:mechanism_overview}.

{\section{Reneging Behavior with Perfect QSI}\label{sec:perfect_qsi}
\subsection{Excluding Reneging from Poisson Queues with Perfect QSI}\label{subsec:poisson_queues_with_perfect_qsi}}
We start our investigation with a trivial case, considering the decision of one arbitrary user under three assumptions:
\begin{enumerate}
	\item the investigated user holds a perfect real-time information of the task queue at MEC server, including its own position in queue and the service rate of MEC server;
	\item tasks are processed by the MEC server as a consistent Poisson process with rate $\mu$;
	\item all users other than the investigated one are patient, or their reneging chances are significantly lower than the service rate, so that the impact of their reneging can be neglected.
\end{enumerate}
When the user generates at $t=0$ a task (but has not made the offloading decision yet), and there are already $k-1$ tasks in the task queue of MEC server, the remaining MEC service latency $\tau\subscript{c}$ in case of entering the queue can be modeled by $\tau_{\mathrm{e},k}=\tau_s+\tau_{\mathrm{w},k}$,
where $\tau_s$ is the overhead latency to offload the task, and $\tau_{\mathrm{w},k}$ is the waiting time at position $k$ in a $G/M/1/\infty$ queue, of which the CDF is independent of $t$:
\begin{equation}
	F_{\tau_{\mathrm{w},k}}(x)=1-\sum_{i=0}^{k-1}\frac{(\mu x)^ie^{-\mu x}}{i!}.
\end{equation}
Thus, $F_{\tau\subscript{c}}(x\vert 0)$ becomes the CDF of $\tau_{\mathrm{e},k}$:
\begin{equation}
	F_{\tau_{\mathrm{c}}}(x\vert0)=F_{\tau_{\mathrm{e},k}}(x)=1-\sum_{i=0}^{k-1}\frac{[\mu (x-\tau\subscript{s})]^ie^{-\mu (x-\tau\subscript{s})}}{i!}.
\end{equation}
On the other hand, for a task that is already offloaded to the MEC server and waiting in the $k\superscript{th}$ position in queue at any $t>0$, its remaining latency is simply $\tau_{\mathrm{w},k}$, and $F_{\tau\subscript{c}}(x\vert 0)$ becomes:
\begin{equation}
	F_{\tau_{\mathrm{c}}}(x\vert t)=F_{\tau_{\mathrm{w},k}}(x)=1-\sum_{i=0}^{k-1}\frac{(\mu x)^ie^{-\mu x}}{i!}, \forall t>0.
\end{equation}
Clearly, with $\tau\subscript{s}\geqslant 0$ there is always $F_{\tau_{\mathrm{e},k}}(x)\leqslant F_{\tau_{\mathrm{w},k}}(x)$, or $F^{-1}_{\tau_{\mathrm{e},k}}(x)\geqslant F^{-1}_{\tau_{\mathrm{w},k}}(x)$, which takes the equality only when $\tau\subscript{s}=0$, so that for all $t\geqslant 0, k\in\mathbb{N}^+, P\subscript{o}\in(0,1)$:
\begin{equation}
	F^{-1}_{\tau\subscript{l}}(1-P\subscript{o})>F^{-1}_{\tau_{\mathrm{e},k}}(1-P\subscript{o})\Rightarrow F^{-1}_{\tau\subscript{l}}(1-P\subscript{o})>F^{-1}_{\tau_{\mathrm{w},k}}(1-P\subscript{o}),
\end{equation}
which implies under the offloading decision rule \eqref{eq:offloading_decision} that
\begin{lemma}\label{lemma:no_rng_after_entrance}
	If a user with perfect information about a $G/M/1/\infty$ FCFS queue does not balk but enters the queue, it does not renege from the queue at its entrance position.
\end{lemma}

Meanwhile, since $F_{\tau_{\mathrm{c}}}(x\vert t)=F_{\tau_{\mathrm{w},k}}(x)$ is independent from $t$, obviously the reneging decision at a certain queue position $k$ does not vary with time, i.e.
\begin{equation}
	\begin{split}
		\exists t_1\in\mathbb{R}^+:& F^{-1}_{\tau\subscript{l}}(1-P\subscript{o})>F^{-1}_{\tau\subscript{c}}(1-P\subscript{o}\vert t_1)\\
		\Rightarrow &F^{-1}_{\tau\subscript{l}}(1-P\subscript{o})>F^{-1}_{\tau\subscript{c}}(1-P\subscript{o}\vert t),\forall t\in\mathbb{R}^+.
	\end{split}
\end{equation}
Furthermore, $F^{-1}_{\tau_{\mathrm{w},k}}(x)$ is strictly monotonically increasing w.r.t. $k\in\mathbb{N}^+$ for all $x\in(0,1)$, i.e. $F^{-1}_{\tau\subscript{l}}(1-P\subscript{o})>F^{-1}_{\tau\subscript{w},k_1}(1-P\subscript{o})\overset{\forall k_1>k_2}{\Longrightarrow} F^{-1}_{\tau\subscript{l}}(1-P\subscript{o})>F^{-1}_{\tau\subscript{w},k_2}(1-P\subscript{o})$. Since the position $k$ of any specific task in a FCFS queue monotonically decreases along with time $t$ in wide sense, i.e. $k(t_1)\geqslant k(t_2)$ for all $t_1<t_2$. Thus, it always holds
\begin{equation}
	\begin{split}
		&F^{-1}_{\tau\subscript{l}}(1-P\subscript{o})>F^{-1}_{\tau\subscript{c}}(1-P\subscript{o}\vert t_1)\\
		\Rightarrow &F^{-1}_{\tau\subscript{l}}(1-P\subscript{o})>F^{-1}_{\tau\subscript{c}}(1-P\subscript{o}\vert t_2), \forall t_1<t_2,
	\end{split}
\end{equation}
which implies
\begin{lemma}\label{lemma:no_regretting_rng}
	If a user with perfect information about a $G/M/1/\infty$ FCFS queue does not renege from the queue at time $t_1$, it does not renege from the queue at any time $t_2>t_1$.
\end{lemma}
Summarizing Lemmas~\ref{lemma:no_rng_after_entrance} and \ref{lemma:no_regretting_rng}, we can conclude that the investigated user will never renege from the queue. Especially, when all users in the queue are fed with perfect QSI, reneging will be eliminated from the system, so that the assumption 3) about all other users being patient applies to every user in the system. Thus, we can conclude that:
\begin{theorem}\label{th:non_regretting}
	Users do not renege from a $G/M/1/\infty$ FCFS queue with its perfect queue status information (QSI) available to all users.
\end{theorem}
This allows us to simplify the mechanism shown in Fig.~\ref{fig:mechanism_overview}: the decision between MEC offloading and local computing only needs to be made once when the task is generated. The reneging behavior is excluded from the system, which is referred to as non-regretting by some literature~\cite{Tirole2010theory}.

Taking a more close view at Theorem~\ref{th:non_regretting}, it defines two conditions for a user with consistent belief $F_{\tau\subscript{l}}$ in local computing delay to be non-regretting in a single-server FCFS queue:
\begin{enumerate}
	\item The service process shall be stationary and Poisson to guarantee a memory-less $F_{\tau_{\mathrm{w},k}}$.
	\item The user shall have perfect QSI, including the Poisson service rate $\mu$ and its real-time position $k(t)$ in the queue.
\end{enumerate}
Unfortunately, both conditions can be usually violated in realistic MEC scenarios.
First, against the assumption of Poisson service, inconsistent service with time-varying rate $\mu_t$ is widely observed in field measurements. Second, making the queue status information fully transparent to all users not only generates an extra signaling overhead, but also may raise concerns about data confidentiality. Towards a queuing mechanism design in realistic scenario, we analyze both cases in the next two sections, respectively.

{\subsection{Risk-based Reneging in Poisson Mixture Queues with Perfect QSI}\label{subsec:poisson_mixture_queues_with_perfect_qsi}}
First we consider the complex non-Poisson queuing process where the instantaneous service rate $\mu_t$ randomly varies over time $t$. This can be generically modeled as a Poisson-mixture process, a.k.a. Markov-modulated Poisson (MMP) process, where at any time $t$, the completion of ongoing service at the server is a Poisson random process, of which the expectation of instantaneous service rate at time $t$ is
\begin{equation}
	\bar\mu_t=\mathbold{\mu}\times \mathbf{y}_t\transpose.
\end{equation}
Here, $\mathbold{\mu}=\left(\mu^{(1)},\mu^{(2)}\dots\mu^{(D)}\right)$ is a $D$-dimensional vector containing the service rate $\mu$ in $D$ different states, and $\mathbf{y}_t=\left(y^{(1)}_t,y^{(2)}_t\dots y^{(D)}_t\right)\transpose$ represents the probability mass function of the server status at time $t$, so that $\vert\mathbf{y}_t\vert=1$ for all $t\in\mathbb{R}^+$. Given the accurate server status observed at any time $t_1$ that $\mu_{t_1}=\mu^{(d)}$, we have $\mathbf{y}_{t_1}=(\underset{d-1}{\underbrace{0,0\dots 0}},1,\underset{D-d}{\underbrace{0,0\dots 0}})$, and the continuous-time Markov behavior of $\mathbf{y}_t$ implies
\begin{equation}
	\mathbf{y}_{t_2}=\mathbf{y}_{t_1}+\int_{t_1}^{t_2}\mathbf{Q}\transpose \mathbf{y}_\tau\diff\tau+\mathbold{m}_{t_2-t_1},
\end{equation}
where $\mathbold{m}_t$ is a $\mathcal{F}^\mathbf{y}$-martingale with $\mathbold{m}_0=\mathbf{0}$, $\mathbf{Q}$ is the inter-state transition rate matrix of the system. With sufficiently large $D$, the MMP model can accurately approximate an arbitrary time-variant service process.

Thus, at any time instant $t$, neglecting the impact of user reneging, the CDF of remaining waiting time at position $k$ becomes~\cite{Ross1996stochastic}
\begin{equation}\label{eq:cdf_remaining_waiting_time}
	F_{\tau_{\mathrm{w},k}}(x\vert t)=1-\sum\limits_{i=1}^{k-1}\frac{\left(\int_{t}^{t+x}\bar\mu_\tau\diff\tau\right)^ie^{-\int_{t}^{t+x}\bar\mu_\tau\diff\tau}}{i!},
\end{equation}
which is jointly determined by $k$ and $\mu_t$. Especially,	if the server status remains consistent during a time interval $[t_1,t_2]$, i.e. $\mu_{t_i}=\mu_{t_j}$ for all $t_i,t_j\in[t_1,t_2]$, the queue is locally $G/M/1/\infty$ during $[t_1,t_2]$, so that the users will not renege, as Theorem~\ref{th:non_regretting} suggests. Applying this on \emph{all} users, the system comes to a equilibrium where \emph{every} user is patient.

When $\mu_t$ is updated, however, $F^{-1}_{\tau_{\mathrm{w},k}}(1-P\subscript{o}\vert t)$ may decrease and a reneging can be triggered%
. Therefore, we propose to update the QSI upon transitions in the service status.

\revise{It is important to remark that upon the updated queue status, all users can be impatient and have chance to renege from the queue, which may reject the assumption of negligible impact by other reneging users on the expected remaining waiting time. More specifically, an increase in $\mu_t$ is suggesting reduced waiting time $\tau_{\mathrm{w},k}$, which will \emph{not} trigger any reneging but only inhibit the balking of new arriving users. To the contrary, a reduction in $\mu_t$ implies longer waiting time $\tau_{\mathrm{w},k}$ for all $k$, and thus generally encouraging the users to renege. The increased reneging rate, however, will reduce the experienced waiting time and thus partially cancel the MEC server's encouragement of reneging. Therefore, the reneging phenomenon will not further excite itself in a closed loop, but eventually achieve a equilibrium. The dynamic process of converging to the new equilibrium depends on the significance of reduction in $\mu_t$.

In case of a slight reduction in $\mu_t$, only a minority of reneging users will be triggered to renege. We have demonstrated in a previous study~\cite{HSC+2020multiservice} of ours, that this phenomenon will not make a significant impact to the overall queue dynamics, and the assumption of $G/M/1/\infty$ queue can still be taken as a good approximation. However, upon  a dramatic drop of the serving rate $\mu_t$ over the queue status transition, a majority of waiting users may be simultaneously triggered to renege from the queue, while \emph{incorrectly} assuming that all other users were patient. In this case, the users' decisions will be significantly biased and very likely non-optimal. 

One possible solution to the issue caused by dense reneging is to correct the CDF model~\eqref{eq:cdf_remaining_waiting_time} regarding the impatience of users. More specifically, when estimating the remaining waiting time of a user at position $k$, the stochastic reneging process of all users waiting ahead of it shall be taken into account. While an analytical solution of this model is hard to obtain, it is possible to rely on the MEC server to empirically learn from its observation to the queue status and therewith online update the CDF model. The updated CDF model, when shared with the users, will also influence the user decisions and therefore change the empirical model again. Therefore, it generally becomes an incomplete information game between the MEC server and the users, which eventually converges to an equilibrium albeit the time it takes.
}

\section{Risk-based Reneging in Poisson Queues with Imperfect QSI}\label{sec:poisson_queues_with_imperfect_qsi}
Another condition that may raise user impatience is the lack of perfect QSI. In a previous work of ours~\cite{HSC+2020multiservice} we have initiated a preliminary research on this case, investigating the impatient queuing behavior regarding an expected profit maximization on different levels of QSI availability. We have learned from the study that a complete blindness of users to the QSI can easily lead to inappropriate decisions of balking/reneging, and therewith impair the system performance. It turns out to be an effective compromise, nevertheless, to feed the users with an imperfect QSI, so as to enable learning-based user decisions while protecting the confidentiality of server status from users.

\subsection{Reneging due to Estimation Errors}\label{subsec:impatience_est_error}
The mechanism of sharing imperfect QSI to users is typically implemented by periodically informing every awaiting user about the position $k$ of its pending task in queue. A common alternative is to inform the user only when $k$ is updated. 
Sometimes, a rough estimation $\mu_0$ can also be provided to all users as an initial belief in the service rate $\mu$, but the accurate value remains unknown and needs to be estimated by the user. Generally, with such mechanisms, a user is able to estimate $\mu$ from its observations on $k$, denoted by $\hat{\mu}(\Omega_t)$, where
\begin{equation}
	\Omega_t=\left(\omega_{t_1},\omega_{t_2},\dots,\omega_{t}\right)
\end{equation}
is the sequence of observation entries at $t$, each entry $\omega_{t_n}=\left(k(t_n),t_n\right)$ consists of the position and the time at observation. A \emph{consistent} estimator $\hat{\mu}$ is guaranteed to converge at $\mu$ in probability:
\begin{align}
	\plim\limits_{t\to+\infty}\hat{\mu}(\Omega_t)=\mu,
\end{align}
where $\plim$ is the probability limit operator.

Nevertheless, within a finite time $t$, despite the consistency of $\hat{\mu}$ there is always a random error between the estimation $\hat{\mu}(\Omega_t)$ and the ground truth $\mu$:
\begin{equation}
	\varepsilon_\mu(\Omega_t)=\hat{\mu}(\Omega_t)-\mu,
\end{equation}
which introduces an error to the decision of reneging. 

More specifically, given the outage risk level $P\subscript{o}$ and the local computing latency distribution $F_{\tau\subscript{l}}(x)$, we recall the offloading decision rule as per Eq.~\eqref{eq:offloading_decision} to see that the correctness of decision relies on the user's belief $F_{\tau\subscript{c}}(x\vert t)$, which is represented in Poisson queues by $F_{\tau\subscript{e},k(t)}(x)$ and determined therefore by the estimation of $\mu$. If a \revise{user} overestimates the service rate $\mu$ with $\varepsilon_\mu(\Omega_0)<0$ when issuing its request at $t=0$, it will also underestimate $F^{-1}_{\tau\subscript{c},k(0)}(1-P\subscript{o})$, and may therewith choose to enter the queue,  although it should have chosen to balk, which we name as \emph{over-patient}; on the other hand, it can also overestimate $F^{-1}_{\tau\subscript{c},k(0)}(1-P\subscript{o})$ with $\varepsilon_\mu(\Omega_0)>0$ and therefore choose to balk, although it should have been non-regretting, which we refer to as \emph{under-patient}.

Such imperfect decisions will, naturally, lead to increased task latency and loss in end utility. A user can rely on nothing but a longer observation sequence $\Omega_t$ to effectively reduce the estimation error, while this learning process itself (implicitly means waiting in the queue) is causing an end-utility decay over leaning time. It becomes therefore a critical task to select an optimal time to stop learning and make the reneging decision, which is discussed in the following.

\subsection{Optimal online Poisson learning: balance between the learning gain and the waiting cost}\label{subsec:optimal_poisson_learning}
To optimize the stopping time of learning process, we need to construct a penalty function to indicate the profit loss caused by inaccurate estimation of $\mu$. As discussed above, there are two classes of error in the decision of reneging, namely over-patient and under-patient. In the earlier case, the \revise{user} underestimates its waiting time in MEC server's queue, and therefore chooses to wait, expecting for an overestimated utility, although the correct decision will reduce the latency by reneging from queue and locally computing instead. In contrast, a \revise{user} in the latter case overestimates the load of MEC server and chooses to compute locally, although a rational decision, i.e. to wait, will lead to a higher utility. Thus, the expected utility loss upon decision error can be correspondingly defined as

\begin{equation}\label{eq:profit_loss}
	\begin{split}
		&L(k,t)=\\
		&\begin{cases}
				u\left(t+\expectation{\tau_1}\right)-\int\limits_{0}^{\mu\subscript{c}(k)}u\left(t+\frac{k}{x}\right)f_{\hat{\mu}}\left(x\vert\Omega_t\right)\diff{x}&\mu>\mu\subscript{c}(k),\\
				\int\limits_{\mu\subscript{c}(k)}^{+\infty}u\left(t+\frac{k}{x}\right)f_{\hat{\mu}}\left(x\vert\Omega_t\right)\diff{x}-u\left(t+\expectation{\tau_1}\right)&\text{otherwise},
			\end{cases}
	\end{split}
\end{equation}
where $\mu_\text{c}(k)$ is the critical arrival rate, which makes $F_{\tau\subscript{w}^{(k)}}(m)=F_{\tau\subscript{l}}(m)$, and $f_{\hat{\mu}}(x\vert\Omega_t)$ is the user's conditional belief about the probability density function (PDF) of estimation $\hat{\mu}$ given observations $\Omega_t$. As the \revise{user} remains waiting in the queue, $\Omega_t$ keeps being updated, and $f_{\hat{\mu}}(x\vert\Omega_t)$ converges to $\delta(x-\mu)$, and $L(k,t)$ also therewith converges to 0.

The stochastic features of $L(k,t)$ is determined by the specific queue status information updating mechanism. As we have named in Section~\ref{subsec:impatience_est_error}, for example, $\Omega_t$ can be updated periodically at every $t=nT$ where $n\in\mathbb{N}$, or updated at every change of $k$ (e.g., when the undergoing service is accomplished at server, or when other users waiting ahead renege from the queue). According to \cite{Zhong2019optimal}, the latter mechanism performs as the optimal dynamic information acquisition for Poisson random processes. Therefore, in this work we consider the queue status information updating upon service accomplishment and user reneging.

Thus, for a certain awaiting request, when its position in queue is updated at $t$ from $k+1$ to $k$, it obtains a marginal learning gain of
\begin{equation}\label{eq:marginal_learning_gain}
	G\subscript{learn}(k,t)=L(k+1,t^-)-L(k,t),
\end{equation}
where $t^-=\lim\limits_{\delta\to 0^+}(t-\delta)$ is the left limit of time instant before the update, so that $\Omega_{t^-}$ does not contain the observation $k(t)$. It is clear that $G\subscript{learn}(k,t)$ monotonically decreases with $t$, converging to 0 as $t\to\infty$, and increases with $k$.

Meanwhile, the cost to further improve the estimate $\hat\mu$ is the expected utility loss until the next update of $k$, i.e. the mean inter-service interval:
\begin{equation}\label{eq:learning_cost}
	C\subscript{learn}(t)=u(t)-u\left(t+\frac{1}{\mu}\right),
\end{equation}
which is always positive and independent of $k$. Thus, we set the optimal decision position at
\begin{equation}
	\begin{split}
		k_\text{opt}(t)=&\sup\left\{k: G\subscript{learn}(k,t)-C\subscript{learn}(t)\leqslant0\right\}\\
		=&\sup\left\{k: G'\subscript{learn}(k,t)-C'\subscript{learn}\leqslant0\right\},
	\end{split}
\end{equation}
where $G'\subscript{learn}(k,t)=G\subscript{learn}(k,t)/u(t)$ and $C'\subscript{learn}(k)=C\subscript{learn}(k,t)/u(t)$.

Note that the ground truth of $\mu$ is essential for the calculation of $L(k,t)$ and $C\subscript{learn}$ according to Eqs.~\eqref{eq:profit_loss} and \eqref{eq:learning_cost}, but unknown to the user. So we propose to use its estimation $\hat{\mu}$ to approximate it. More specifically, we apply the bias-corrected maximum likelihood estimator
\begin{equation}\label{eq:mu_est}
	\hat{\mu}(\Omega_t)=\begin{cases}
		+\infty&N_t=1,\\
		\frac{N_t-1}{\sum\limits_{n=2}^{N_t}\Delta t_n}&N_t=2,\\
		\frac{\left(N_t-1\right)\left(N_t-3\right)}{\left(N_t-2\right)\sum\limits_{n=2}^{N_t}\Delta t_n}&N_t\geqslant 3,
	\end{cases}
\end{equation}
where $N_t=\left\vert\Omega_t\right\vert_0$  is the number of observations at $t$, and $\Delta t_n=t_n-t_{n-1}$ denotes the $\left(n-1\right)^\text{th}$ inter-update interval. Remark that since 
\begin{equation}
	\Delta t_n\sim\text{Exp}\left(\frac{1}{\mu}\right),\quad\forall n\geqslant 2,
\end{equation}
we have 
\begin{equation}
	\sum\limits_{n=2}^{N_t}\Delta t_n\sim\text{Erlang}\left(N_t-1,\frac{1}{\mu}\right),\quad\forall N_t\geqslant 2,
\end{equation}
which can be exploited to calculate $f_{\hat{\mu}}(x\vert\Omega_t)$.

Our proposed risk-based dynamic reneging mechanism with online learning can be briefly described by Algorithm~\ref{alg:reneging}.

	\begin{algorithm}[!htbp]
		\caption{The risk-based dynamic user reneging algorithm}
		\label{alg:reneging}
		\footnotesize
		\DontPrintSemicolon
		Initialization (queue entrance at $k$): $n=0, \Omega=\emptyset, k, P\subscript{o}, F_{\tau\subscript{l}}, \hat{\mu}=+\infty, R=\text{False}$\;
		\While(\hfill\emph{Wait}){$R=\text{False}$}{
			Wait for the next observation $\omega_{n+1}$ or service\;
			$n\gets n+1$\;
			\If{Served}{\Return{served}}
			$k\gets k-1$\;
			$\Omega\gets\Omega\cup\{\omega_n\}$\;
			Update $F_{\tau_{\mathrm{w},k}}$ and $\hat{\mu}$\;
			\If{$G'\subscript{learn}(k,t)\leqslant C'\subscript{learn}(k) \And F^{-1}_{\tau\subscript{l}}(1-P\subscript{o})\leqslant F^{-1}_{\tau_{\mathrm{w},k}}(1-P\subscript{o})$}{
				\Return{reneged}
			}
		}
		\Return{reneged}
	\end{algorithm}

\section{Numerical Evaluation}\label{sec:sim}
\subsection{Simulation Setup}
To verify and evaluate our proposed approaches, an exhaustive simulation campaign is conducted, in order to benchmark the risk-based reneging mechanism with classical baseline methods in various service scenarios and under different information conditions. 

In the simulations, we consider an exponentially decaying utility
\begin{equation}
	u(\Delta t)=u_0e^{-\beta\Delta t}
\end{equation}
where $\beta$ is the discount exponent. We consider consistent Poisson arrivals of new requests at the service queue. Every request is assigned with an exponential distribution of the local computation latency, for which the mean value is individually randomly generated for each request. For the MEC service, we ignore the small delay caused by offloading overhead, and consider two states of the server to represent different load levels, each dedicated to a specific service rate. Three different MMP scenarios are specified, each characterized by a inter-state transfer rate matrix that represents the dynamics of transition between the two aforementioned server states. Detailed specifications are listed in Tab.~\ref{tab:sim_setup}.
\begin{table}[!hbtp]
	\centering
	\begin{tabular}{c|c|c|l}
		\toprule[2px]
		\multicolumn{2}{c|}{\textbf{Parameter}}			&\textbf{Value}				&\textbf{Description}\\\midrule[1.5px]
		\multicolumn{2}{c|}{$\lambda$}					&1.5						&Request arrival rate\\\hline
		\multirow{2}{*}{$\bar\tau_\text{l}$}			
								&Model 1:				&$\sim\mathcal{U}(2,10)$	&\multirow{2}{*}{Mean local computation latency}\\
								&Model 2:				&$\sim\mathcal{U}(4,15)$	&\\\hline
		\multicolumn{2}{c|}{$\tau_\text{s}$}			&0							&Task offloading overhead latency\\\hline
		\multicolumn{2}{c|}{$\mu_1$}					&2							&Service rate (low server load)\\\hline
		\multicolumn{2}{c|}{$\mu_2$}					&1							&Service rate (high server load)\\\hline
								&Scenario A: 			&$\begin{pmatrix}
								 							0.9&0.1\\
															0.8&0.2
															\end{pmatrix}$			&\\
		$\mathbf{Q}$			&Scenario B:			&$\begin{pmatrix}
															0.7&0.3\\
															0.3&0.7
															\end{pmatrix}$			&Inter-state transition rate matrix\\
								&Scenario C: 			&$\begin{pmatrix}
															0.2&0.8\\
															0.1&0.9
															\end{pmatrix}$			&\\\hline
		\multicolumn{2}{c|}{$u_0$}						&1							&Initial utility\\\hline
		\multicolumn{2}{c|}{$\beta$}					&0.1						&Discount exponent\\\hline
		\multicolumn{2}{c|}{$T_0$}						&1							&Simulation time resolution\\
		\bottomrule[2px]
	\end{tabular}
	\caption{Simulation setup}
	\label{tab:sim_setup}
\end{table}

{Note that according to our analysis at the beginning of Section~\ref{subsec:poisson_queues_with_perfect_qsi} that led us to Lemma~\ref{lemma:no_rng_after_entrance}, the analytical structure of the user behavior (i.e., not reneging from the queue at entrance) is valid regardless of the exact value of offloading overhead $\tau\subscript{s}$, as long as $\tau\subscript{s}\geqslant 0$. Any positive offloading overload, either random or deterministic, will shift the user belief in waiting time $\tau\subscript{c}$ towards higher value, while leaving the belief in $\tau\subscript{l}$ uninfluenced. Thus, a user is more likely to balk with a higher offloading overhead than with a lower one, if all other conditions remain the same. However, this is only a quantitative difference that makes no structural change in the user behavior. Since it is not in the scope of this work to evaluate the practical performance under realistic specification of traffic and business models, here we choose to omit the impact of $\tau\subscript{s}$, setting it to zero for a simple implementation and a better focus on the analytical structure. }

\subsection{Poisson Service with Perfect QSI}\label{subsec:sim_perfect_poisson_qsi}
\begin{figure}[!hbtp]
	\centering
	\begin{subfigure}[t]{\linewidth}
		\centering
		\includegraphics[width=.7\linewidth]{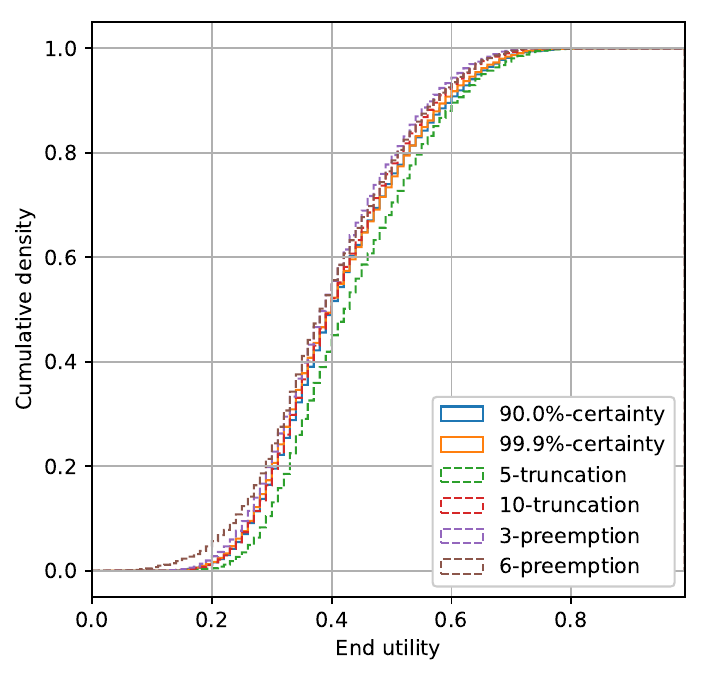}
		\caption{Low server load, local latency model 1}
	\end{subfigure}	
	\begin{subfigure}[t]{\linewidth}
		\centering
		\includegraphics[width=.7\linewidth]{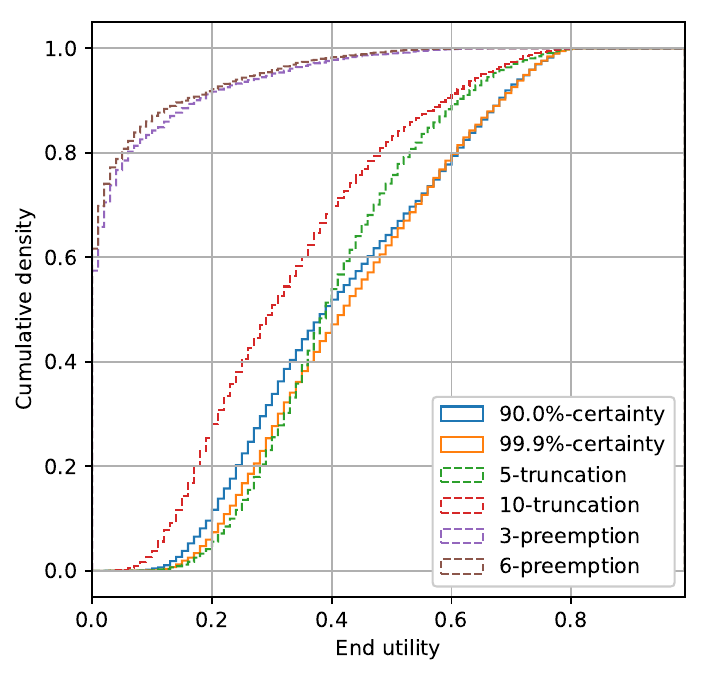}
		\caption{High server load, local latency model 1}
	\end{subfigure}
	\caption{End utility distribution of Poisson queue congestion control approaches with perfect QSI}
	\label{fig:benchmark_perfect_poisson_qsi}	
\end{figure}

We started with the trivial case there the service is Poisson and a perfect QSI is available for all \revise{users}, and conducted the simulation in both the service scenarios with high and low server load. In each scenario, we tested different queue congestion control approaches under variate specifications, including the risk-based reneging method with outage risk level $P_\mathrm{o}\in\{0.1,0.001\}$, the queue truncation method with maximal queue length ${l}_\mathrm{max}\in\{5,10\}$, and the preemptive server method with preemption timeout $\tau_{\max}\in\{3,6\}$. {More specifically, with the queue truncation method, if the instantaneous queue length $l<l_\mathrm{max}$, any arriving task will always be offloaded to the MEC server and patiently waits in the queue; when $l=l_\mathrm{max}$, any arriving task will be processed locally. With the preemption method, every task can be only processed by the MEC server for no longer than $\tau_{\max}$, after that timeout it will be automatically removed from the server and sent back to local processing.} The local computation latency model was set to model 1, i.e. $\bar\tau_\text{l}\sim\mathcal{U}(2,10)$. Every specific solution was evaluated through 10 independent Monte-Carlo tests, each lasting $300$ periods, and the performance is measured by CDF of end utility, request admission rate, average end utility, and median end utility. The results are briefly summarized in Fig.~\ref{fig:benchmark_perfect_poisson_qsi} and Tab.~\ref{tab:benchmark_perfect_poisson_qsi}.
\begin{table}[!hbtp]
	\centering
	\caption{KPIs of Poisson queue congestion control approaches with perfect QSI}
	\label{tab:benchmark_perfect_poisson_qsi}
	\begin{tabular}{lc|c c c|c c c}
		\toprule[2px]		
		&			&\multicolumn{3}{c|}{\textbf{Low server load}} 	&\multicolumn{3}{c}{\textbf{High server load}}\\
		&			&\multicolumn{3}{c|}{\textbf{Local latency model 1}} 	&\multicolumn{3}{c}{\textbf{Local latency model 1}}
		\\\midrule[1px]
		\multicolumn{2}{c|}{\textbf{Method}}&	Adm		&	Avg		& 	Med		&	Adm		&	Avg		& 	Med		
		\\\hline
		\multirow{2}{*}{R}		&	0.1\%	&	97.57\%	&	0.4192	&	0.4018	&	58.26\%	&	0.4414	&	0.4250
		\\
								&	10\%	&	97.18\%	&	0.4222	&	0.4029	&	62.67\%	&	0.4242	&	0.3957
		\\\hline
		\multirow{2}{*}{T}		&	5		&	89.34\%	&	0.4432	&	0.4295	&	61.43\%	&	0.4134	&	0.3957	
		\\
								&	10		&	98,31\%	&	0.4159	&	0.4029	&	66.60\%	&	0.3357	&	0.3050	
		\\\hline
		\multirow{2}{*}{P}		&	3		&	82.91\%	&	0.4047	&	0.3906	&	74.82\%	&	0.0525	&	0.0049	
		\\
								&	6		&	90.72\%	&	0.4018	&	0.3891	&	85.91\%	&	0.0458	&	0.0041
		\\\midrule[1px]
		\multicolumn{8}{l}{Methods: risk-based \underline{R}eneging; queue \underline{T}runcation; server \underline{P}reemption}
		\\
		\multicolumn{8}{l}{KPIs: \underline{Adm}ission rate; \underline{Av}era\underline{g}e end utility; \underline{Med}ian end utility}
		\\\bottomrule[2px]
	\end{tabular}
\end{table}

Generally, we can see that all methods work well when the server is under-loaded, i.e. when the service rate is higher than the arrival rate. However, when the server is overloaded, i.e. when the service rate drops below the arrival rate, both the baseline methods of queue truncation and server preemption are significantly disturbed and suffer from reduced utilities. In contrast, the risk-based reneging approach outperforms the baselines with a good robustness against the congestion: it delivers not only barely influenced average/median utility, but also much heavier a tail regarding the baselines in utility distribution. In other words, compared to the baselines, the risk-based reneging method exhibits more preference to the peak-performance regime than the mid-performance regime, while generally enhancing the average performance. Furthermore, by comparing the CDF curves of $90.0\%$-certainty-reneging and $99.9\%$-certainty-reneging, we can observe that setting a higher risk level $P_\mathrm{o}$ leads to a smoothed probability density of end utility (i.e. both the chances of getting high utility and low utility increase, with the chance of having intermediate utility reduced).


\subsection{MMP Service with Perfect QSI}\label{subsec:sim_perfect_mmp_qsi}
\begin{figure*}[!hbtp]
	\centering
	\begin{subfigure}[t]{.49\linewidth}
		\centering
		\includegraphics[width=.7\linewidth]{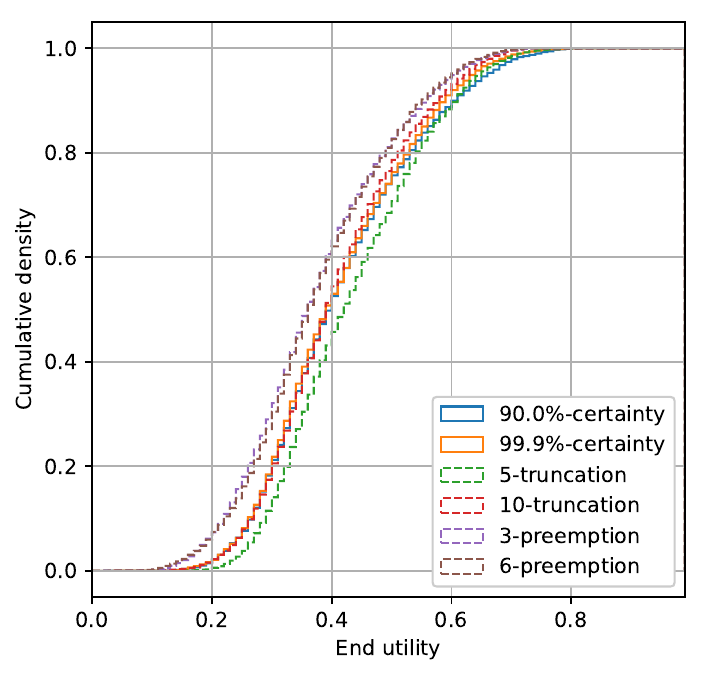}
		\caption{Scenario A, local latency model 1}
	\end{subfigure}	
	\begin{subfigure}[t]{.49\linewidth}
		\centering
		\includegraphics[width=.7\linewidth]{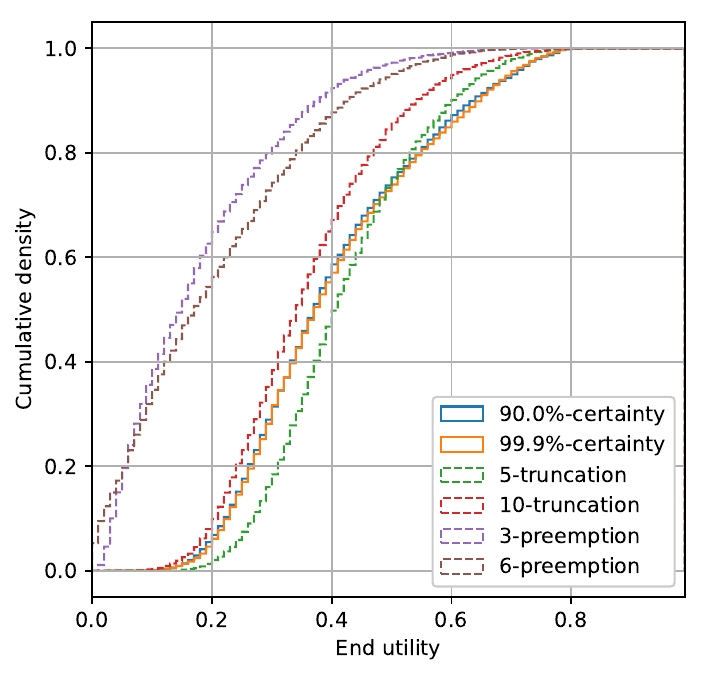}
		\caption{Scenario B, local latency model 1}
	\end{subfigure}	
	\begin{subfigure}[t]{.49\linewidth}
		\centering
		\includegraphics[width=.7\linewidth]{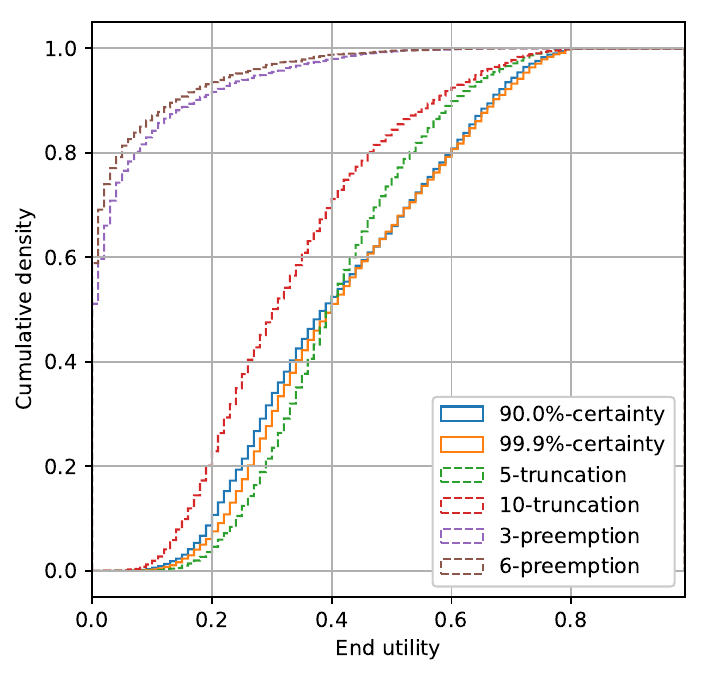}
		\caption{Scenario C, local latency model 1}
	\end{subfigure}
	\begin{subfigure}[t]{.49\linewidth}
		\centering
		\includegraphics[width=.7\linewidth]{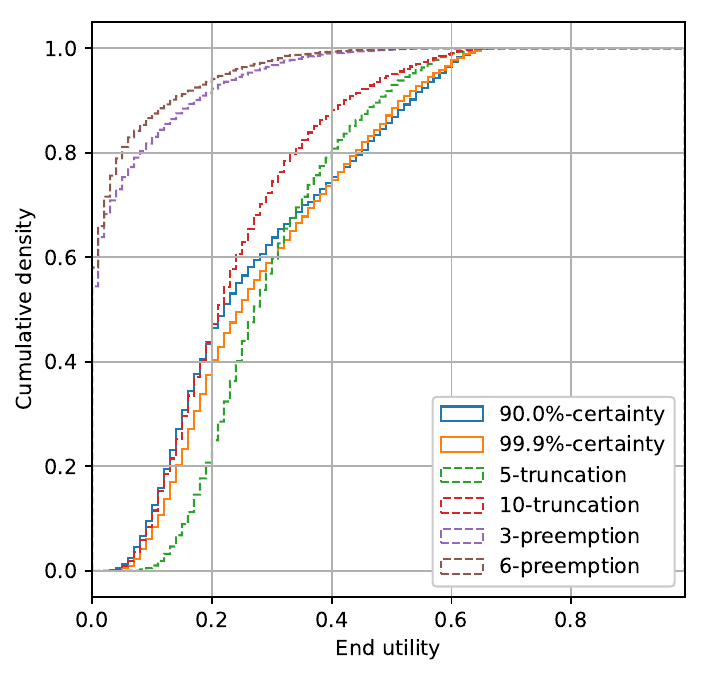}
		\caption{Scenario C, local latency model 2}
	\end{subfigure}	
	\caption{End utility distribution of MMP queue congestion control approaches with perfect QSI}
	\label{fig:benchmark_perfect_mmp_qsi}
\end{figure*}

Then we investigated the case of non-Poisson service, where the server status randomly switches between \revise{unburdened} and \revise{overloaded} as a Markov process. Three such MMP scenarios A--C were defined as described in Tab.~\ref{tab:sim_setup}, and in each scenario simulations were conducted with local latency specified to model 1, in the same way as in Section~\ref{subsec:sim_perfect_poisson_qsi}. Additionally, to evaluate the sensitivity to local latency, we also tested a combination of MMP scenario C and local latency model 2 where $\bar\tau_\text{l}\sim\mathcal{U}(4,15)$.
The general results are, as summarized by Fig.~\ref{fig:benchmark_perfect_mmp_qsi} and Tab.~\ref{tab:benchmark_perfect_mmp_qsi}, also similar to the Poisson service case: the proposed risk-based reneging mechanism works similarly well as the baselines under moderate queue congestion (Scenario A), and gradually stands out when the congestion becomes denser (Scenario B and C). Its extra preference in peak-utility regime over mid-utility regime remains also significant here. Without any surprise, the risk-based reneging method shows a higher sensitivity to local computation latency than the baselines, tending to offload more tasks to the MEC when the local computation is slower.

\begin{table*}[!hbtp]
	\scriptsize
	\centering
	\caption{KPIs of MMP queue congestion control with perfect QSI, local computation latency specified to Model 1}
	\label{tab:benchmark_perfect_mmp_qsi}
	\begin{tabular}{lc|c c c|c c c|c c c|c c c}
		\toprule[2px]		
									&			&\multicolumn{3}{c|}{\textbf{Scenario A}} 	&\multicolumn{3}{c|}{\textbf{Scenario B}} 	&\multicolumn{6}{c}{\textbf{Scenario C}}
		\\\cline{9-14}
									&			&\multicolumn{3}{c|}{\textbf{Local latency model 1}} 	&\multicolumn{3}{c|}{\textbf{Local latency model 1}} 	&\multicolumn{3}{c|}{\textbf{Local latency model 1}}	&\multicolumn{3}{c}{\textbf{Local latency model 2}}
		\\\midrule[1px]
		\multicolumn{2}{c|}{\textbf{Method}}	&	Adm		&	Avg		& 	Med		&	Adm		&	Avg		& 	Med		&	Adm		&	Avg		& 	Med		&	Adm		&	Avg		& 	Med	  \\\hline
		\multirow{2}{*}{R}			&	0.1\%	&	96.94\%	&	0.4156	&	0.3971	&	82.66\%	&	0.4094	&	0.3781	&	64.85\%	&	0.4302	&	0.4043	&	70.41\%	&	0.2905	&	0.2519\\
									&	10\%	&	94.57\%	&	0.4212	&	0.4008	&	84.33\%	&	0.4059	&	0.3764	&	66.74\%	&	0.4207	&	0.3913	&	71.27\%	&	0.2782	&	0.2255\\\hline
		\multirow{2}{*}{T}			&	5		&	87.29\%	&	0.4395	&	0.4252	&	78.41\%	&	0.4267	&	0.4108	&	66.05\%	&	0.4170	&	0.4015	&	65.30\%	&	0.3015	&	0.2785\\
									&	10		&	97,43\%	&	0.4119	&	0.3965	&	89.46\%	&	0.3645	&	0.3471	&	72.86\%	&	0.3395	&	0.3097	&	72.93\%	&	0.2443	&	0.2176\\\hline
		\multirow{2}{*}{P}			&	3		&	82.76\%	&	0.3799	&	0.3641	&	81.43\%	&	0.1846	&	0.1515	&	76.81\%	&	0.0546	&	0.0087	&	77.16\%	&	0.0513	&	0.0070\\
									&	6		&	90.55\%	&	0.3816	&	0.3668	&	89.73\%	&	0.2076	&	0.1764	&	87.01\%	&	0.0435	&	0.0055	&	87.01\%	&	0.0411	&	0.0051	\\\bottomrule[2px]
	\end{tabular}
\end{table*}
%

\subsection{Poisson Service with Imperfect QSI}\label{subsec:sim_imperfect_poisson_qsi}
\begin{figure}[!hbtp]
	\centering
	\begin{subfigure}[t]{\linewidth}
		\centering
		\includegraphics[width=.7\linewidth]{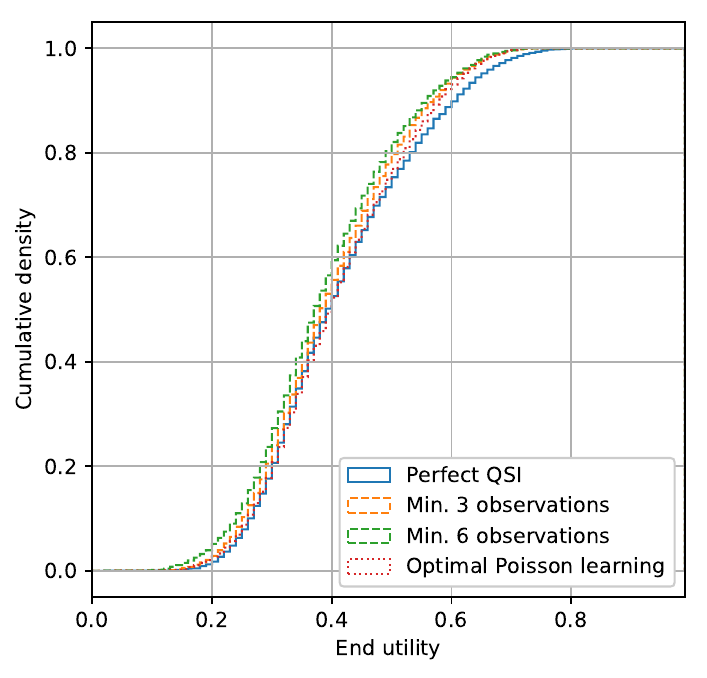}
		\caption{Low server load, local latency model 1}
	\end{subfigure}	
	\begin{subfigure}[t]{\linewidth}
		\centering
		\includegraphics[width=.7\linewidth]{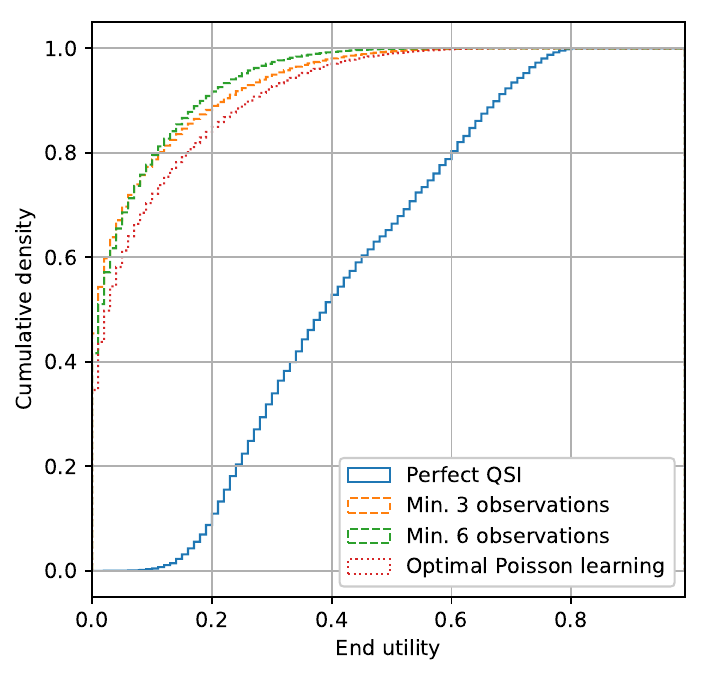}
		\caption{High server load, local latency model 1}
	\end{subfigure}
	\begin{subfigure}[t]{\linewidth}
		\centering
		\includegraphics[width=.7\linewidth]{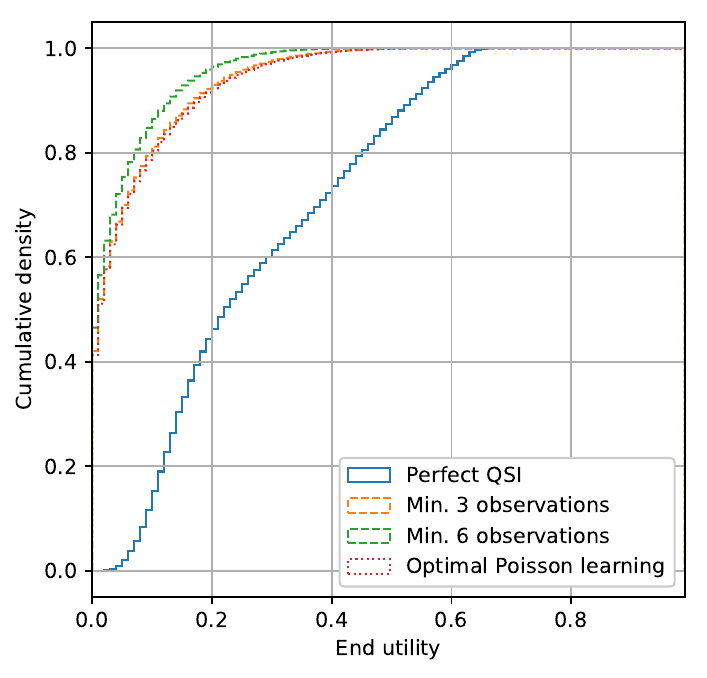}
		\caption{High server load, local latency model 2}
	\end{subfigure}
	\caption{End utility distribution of Poisson queue congestion control with imperfect QSI}
	\label{fig:benchmark_imperfect_poisson_qsi}
\end{figure}

Regarding the case of Poisson service with imperfect QSI, we tested the optimal Poisson learning mechanism proposed in Section~\ref{subsec:optimal_poisson_learning}. For a benchmark we also tested two static learning policies as baselines, which take at least 3 and 6 observations before considering reneging, respectively. In addition, reneging with perfect QSI is also measured as a reference for upper bound performance. Once again, the benchmark was carried out under both server load scenarios with local latency model 1, each through 10 independent 300-period Monte-Carlo tests. An additional combination of high server load and local latency model 2 was tested to evaluate the local latency sensitivity. The results are summarized in Fig.~\ref{fig:benchmark_imperfect_poisson_qsi} and Tab.~\ref{tab:benchmark_imperfect_poisson_qsi}. It can be observed that the loss caused by improper reneging decision based on imperfect QSI is negligible when the server is under-loaded, but dramatically increases under a dense queue congestion, regardless of the learning strategy. A high sensitivity to the local latency can also be generally observed, independent from the learning strategy. Nevertheless, the optimal Poisson learning policy manages to adjust its learning phase regarding the queue status, and therewith obtains a significant utility gain over the static learning (up to $30\%$ in average utility and $63\%$ in median utility).

\begin{table*}[!hbtp]
	\centering
	\caption{KPIs of Poisson queue congestion control with imperfect QSI}
	\label{tab:benchmark_imperfect_poisson_qsi}
	\begin{tabular}{lc|c c c|c c c|ccc}
		\toprule[2px]		
		&			&\multicolumn{3}{c|}{\textbf{Low server load}} 			&\multicolumn{6}{c}{\textbf{High server load}}\\\cline{6-11}
		&			&\multicolumn{3}{c|}{\textbf{Local latency model 1}} 	&\multicolumn{3}{c|}{\textbf{Local latency model 1}}	&\multicolumn{3}{c}{\textbf{Local latency model 2}}
		\\\midrule[1px]
		\multicolumn{2}{c|}{\textbf{Method}}				&	Adm		&	Avg		& 	Med		&	Adm		&	Avg		& 	Med		&	Adm		&	Avg		& 	Med	\\\hline
		\multicolumn{2}{l|}{Optimal Poisson learning}		&	91.91\%	&	0.4144	&	0.4025	&	1.51\%	&	0.0874	&	0.0304	&	1.53\%	&	0.0594	&	0.0186	\\\hline
		\multirow{2}{*}{Fixed minimal observations}	&	3	&	93.31\%	&	0.4040	&	0.3894	&	1.24\%	&	0.0671	&	0.0144	&	1.52\%	&	0.0571	&	0.0175	\\
													&	6	&	94.68\%	&	0.3919	&	0.3772	&	1.52\%	&	0.0610	&	0.0186	&	1.59\%	&	0.0433	&	0.0129	\\\hline
		\multirow{2}{*}{Perfect QSI}				&		&	96.00\%	&	0.4204	&	0.3992	&	62.23\%	&	0.4211	&	0.3926	&	64.92\%	&	0.2795	&	0.2275	\\
													&		&\multicolumn{3}{c|}{(Average 1.612 obs)}		&	\multicolumn{3}{c|}{(Average 2.530 obs)}	&	\multicolumn{3}{c}{(Average 2.545 obs)}
		\\\bottomrule[2px]
	\end{tabular}
\end{table*}

{\section{Further Discussions}
\subsection{Inter-Task Dependency}\label{subsec:inter-task_dep}
For the convenient of model setup and analysis, in this work we have generally assumed independent arrivals and services of individual tasks. Nevertheless, inter-task dependency will not invalidate our proposed approach, as detailed below in three aspects:

\begin{enumerate}
	\item Inter-task arrival dependency: this will be reflected in non-Poisson arrivals of new tasks at the service queue. In FCFS queues, it is clear that the decision about impatient behavior by every \revise{user} depends only on the previous tasks that had arrived before its own task, so the arrival pattern of new tasks will not make any impact on such decisions.
	
	\item Server-state-dependent service: this will be reflected in a time-varying service rate that is determined by the server state, which can be captured by invoking the MMP model. Depending on the mechanism implementation, the instantaneous service rate may be either perfectly or only partially observable to \revise{users}.  The earlier case has been fully discussed in Sections~\ref{subsec:poisson_mixture_queues_with_perfect_qsi} and \ref{subsec:sim_perfect_mmp_qsi}; the latter can only be resolved by an efficient online learning strategy to estimate non-Poisson processes with generically distributed queuing delay, which we identify as a direction of future study.
	
	\item Heterogeneous services: this will be reflected in a service rate that is dependent on the service type  of each individual task. \begin{enumerate}
		\item In a single FCFS queue, the time-varying service rate at server is solely determined by the sequence of tasks, i.e. dependent on the random arrival sequence of heterogeneous tasks. While every \revise{user} is aware of the type of its own task, it is not supposed to know the type of other tasks, and the time-varying service rate to every \revise{user} can be also equivalently modeled with the MMP model. Indeed, each \revise{user} is able to exploit the additional knowledge of its own task type to obtain a moderate gain by knowing the service rate it experiences at the server. This enhancement, however, is straightforward and not focused on by this work. 
		\item Another possible implementation is to set up multiple queues, each only serving tasks of one specific service type, and the computing resource is shared among different queues upon the service provider's policy, as applied in our previous work~\cite{HSC+2020multiservice}. In this case, the services for different queues are independent from each other under any certain resource allocation among queues, and each queue can be separately analyzed under the assumption of independent service among the tasks it contains.
	\end{enumerate}
\end{enumerate}

\subsection{Meeting Specific Latency Requirement}\label{subsec:latency_req} 
Another challenge that can be raised by heterogeneous services is to meet certain strict latency requirements for specific tasks that are extremely sensitive to delay, e.g. for ultra-reliable low-latency communication (URLLC) services. Our proposal in this work, indeed, does not address this issue by itself, because it does not aim at granting any certain task (nor all tasks in average) with a guaranteed quality of service over some specific lower bound. Instead, it guarantees that every individual task takes the \emph{better} option between MEC offloading and local computing with a certainty in the decision making. If both options are promising only poor performance, e.g. when the local device has low computing power and the MEC server is also overloaded, our proposed approach cannot solely guarantee to deliver the required performance, but only help every task reduce the expected latency by taking the option that is more likely better than the other.

However, tasks generally differ from each other in \emph{i}) the belief of local computing latency due to different local computing power, and \emph{ii}) the belief of waiting time due to the difference position in queue. Hence, the tasks that will more likely profit from MEC offloading than local computing will tend more to stay in the queue, while the ones that are more likely to suffer from waiting latency will tend more to balk/renege. Thus, overall, the approach is capable of reducing the latency, raising the utility rate and improving the MEC efficiency in a statistical manner, as we have observed from the numerical results.

Nevertheless, in practical deployment where heterogeneous tasks in various service classes coexist and share the same MEC server, it is important to discriminate latency-\revise{tolerant} tasks against latency-\revise{sensitive} ones. To realize this, the aforementioned multi-queuing mechanism~\cite{HSC+2020multiservice} can be introduced, where tasks are assigned to different queues regarding the service class. \revise{More specifically, the computing resource on MEC server shall be intelligently and flexibly allocated among the queues, so that the latency-sensitive queues are guaranteed with different certain expectations of waiting time regarding their priorities, and the remaining resource allocated to the latency-tolerant queue to provide an elastic service. Thus, at the the same position in queue, tasks in more prioritized queues have more left-skewed CDF of remaining waiting time than those in less prioritized queues, so that different queues are generally discriminated against each other. On top of this multi-queuing framework, our} proposed risk-based reneging mechanism can be independently applied on every individual queue to \revise{further} improve the intra-queue performance. 

\subsection{Multi-Objective Optimization and \revise{User} Heterogeneity}\label{subsec:het_user}
It shall be remarked that in practice, heterogeneity exists not only in the service type that has been above-mentioned, but also in numerous other parameters such as device type, user mobility, energy constraint, etc. Towards a real deployment of our proposed solution in practical MEC scenario, these parameters have to be considered by each individual \revise{user} to establish its own utility function and the corresponding statistical model. Nevertheless, focusing on demonstrating the technical potentials of exploiting impatient queuing behaviors in MEC, we have omitted these details to keep the study in a generic fashion. Since the proposed decision making mechanism and framework do not rely on any specific environment or device condition, but only on the \revise{monotone} of utility w.r.t. latency, their applicability will not be rejected by the practical scenario.

\revise{
\subsection{Multi-Server Scenario}
While this article is focusing on a simplified case where the users decides between the options of computing the task locally and offloading it to \emph{the only} MEC server, in practical scenario, there may be multiple MEC servers available for the users to choose. The policy to select the MEC server for initial offload of the task can be obtained by straightforwardly extending Eq.~\ref{eq:offloading_decision}: the user shall take into account the believes of waiting time at \emph{all} MEC servers as well as local computing, and take the option with lowest outage risk. However, it further introduces an additional challenge, when the users want to flexibly switch between the queues of different MEC servers regarding their instantaneous QSI. Unlike the balking/reneging behavior investigated in this article, the switching mechanism is symmetric, i.e. a user shall be allowed to go back and forth between different queues when necessary. Therefore, such a phenomenon shall be considered as another type of impatient queuing behavior: the \emph{jockeying}~\cite{koenigsberg1966jockeying}.

Albeit the rich literature from the field of queuing theory that model and analyze the jockeying phenomenon, there has been little effort made to investigate its application and design in the context of MEC, to the best of our knowledge. Besides, existing jockeying methods, such as \cite{dehghanian2016strategic}, generally aim at minimizing the expected waiting cost of users. By introducing multiple servers to the problem investigated in this article and allowing symmetric switching between servers, it can be straightforwardly extended into a possible novel design of risk-based jockeying mechanism.
}
}

\section{Conclusion and Outlooks}\label{sec:conclusion}
In this work we have studied the risk-based \revise{user} reneging mechanism, which appears a promising decentralized solution to control the MEC queue congestion with sensitivity to latency outage risk. For generic MEC tasks with time-sensitive utility, we have investigated the rational \revise{user} choice between waiting within the MEC task queue and reneging for local computation, and identified the conditions of balking and reneging in Poisson service queues. 

Our analysis proves that a time consistency of risk preference can sufficiently eliminate the reneging phenomenon{,} when a perfect {statistical} knowledge about queue status is available a-priori to the \revise{users}. Then, we have investigated two cases where the non-regretting \revise{user} behavior is violated and \revise{user} reneging is activated: the non-Poisson service time and the imperfect queue status information. For non-Poisson services, we have introduced the MMP model to generalize the impatience-based \revise{user} decision mechanism. Regarding the case of imperfect queue status information, we have proposed an optimal online Poisson learning strategy that balances between learning gain and waiting loss. Our proposed approaches have been verified effective in various service scenarios through exhaustive numerical simulations. \revise{The proposed novel approach allows users with computing tasks to rationally choose between MEC offloading and local computing regarding its own risk of latency outage, without having to provide its latency or utility model to the MEC service provider. It can be used as a competitive solution of decentralized queue management for multi-user scenarios that are intolerant to latency and privacy, such as autonomous driving, AI-as-a-Service, and IoT over public mobile network.}

While we are focusing in this study on the utility loss caused by latency, the energy consumption of user devices also plays a key role in the practical applications of MEC task offloading. It calls for a significant follow-up effort to design a multi-objective MEC queue management framework, which supports \revise{user} impatience while taking into account of the user mobility and channel condition. Besides, though we have addressed the reneging phenomenon for both scenarios of non-Poisson service and imperfect Poisson QSI, respectively, the impatience-based approach is still challenged by a more complicated case, where imperfect QSI and non-Poisson service time are combined. An efficient online learning strategy to estimate a generic distribution of queuing delay{, as we have discussed in Section~\ref{subsec:het_user},} is therefore required. \revise{Upon dense reneging that violates the assumption of $G/M/1/\infty$ queue, the incomplete information game between the MEC server and users to learn the effective queuing model online and converge to an equilibrium, which is discussed in Sec.~\ref{subsec:poisson_mixture_queues_with_perfect_qsi}, is worth of investigation.} Additionally, in specific and practical use scenarios where heterogeneous services coexist and share the same MEC server, or when users are allowed to freely choose and flexibly switch among multiple MEC servers, it is of a great interest to evaluate the end performance of our proposed approach in combination with a multi-queuing mechanism. \revise{Furthermore, an multi-server extension that considers risk-based jockeying among different MEC servers is also a possible direction of future study.}

\revise{\section*{Acknowledgment}
The work of Technical University of Kaiserslautern was partly supported by the German Federal Ministry of Education and Research (BMBF) within the project Open6GHub under grant number 16KISK004. The work of NEC Laboratories Europe was supported by the European Union's Horizon 2020 project RISE-6G under grant number 101017011. The work of University of Lausanne was supported by Swiss National Science Foundation (SNSF) under grant number 100018\_192583}

  \newpage



%

\newpage


\begin{IEEEbiography}[{\includegraphics[width=1in,height=1.25in,clip,keepaspectratio]{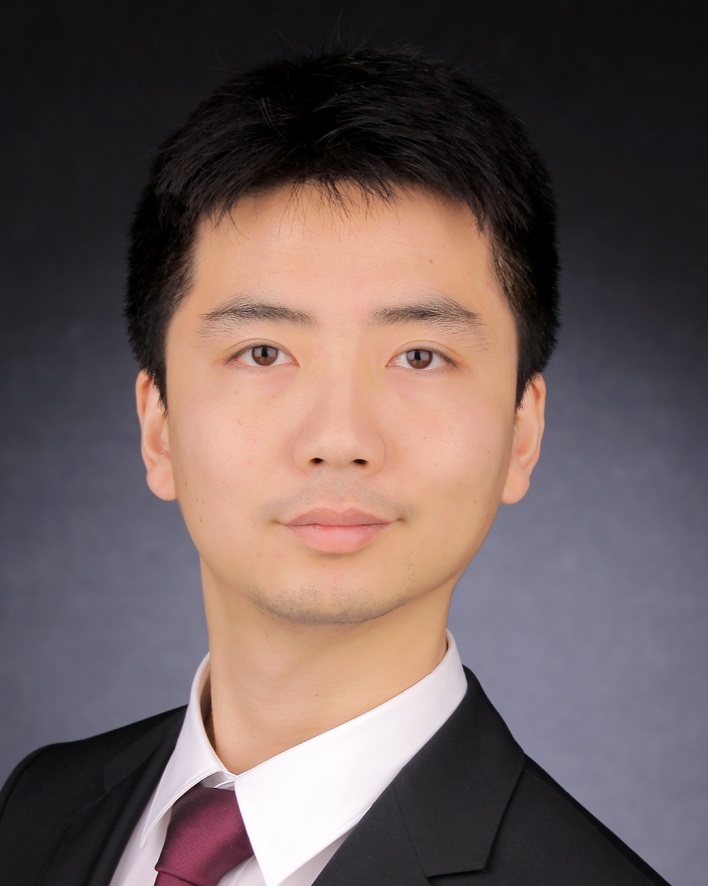}}]{Bin Han} (M'15--SM'21)
	received his B.E. degree in 2009 from Shanghai Jiao Tong University, M.Sc. in 2012 from Technical University of Darmstadt, and the Ph.D. degree in 2016 from Karlsruhe Institute of Technology. Since July 2016 he has been with Technical University of Kaiserslautern as Postdoctoral Researcher and Senior Lecturer. His research interests are in the broad area of wireless communication and networking, with current focus on B5G/6G and MEC. He is the author of over 50 research papers and book chapters, and has participated in multiple EU Horizon 2020 projects. He serves as Guest Editor of \emph{Network} and \emph{Electronics}. 
\end{IEEEbiography}

\begin{IEEEbiography}[{\includegraphics[width=1in,height=1.25in,clip,keepaspectratio]{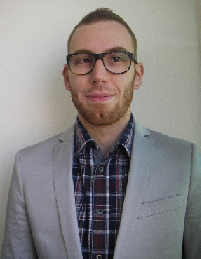}}]{Vincenzo Sciancalepore} (S'11--M'15--SM'19) received his M.Sc. degree in Telecommunications Engineering and Telematics Engineering in 2011 and 2012, respectively, whereas in 2015, he received a double Ph.D. degree. Currently, he is a Principal Researcher at NEC Laboratories Europe in Heidelberg, focusing his activity on network virtualization and network slicing challenges. He is the Chair of the ComSoc Emerging Technologies Initiative (ETI) on Reconfigurable Intelligent Surfaces (RIS) and an editor of IEEE Transactions on Wireless Communications.
\end{IEEEbiography}

\begin{IEEEbiography}[{\includegraphics[width=1in,height=1.25in,clip,keepaspectratio]{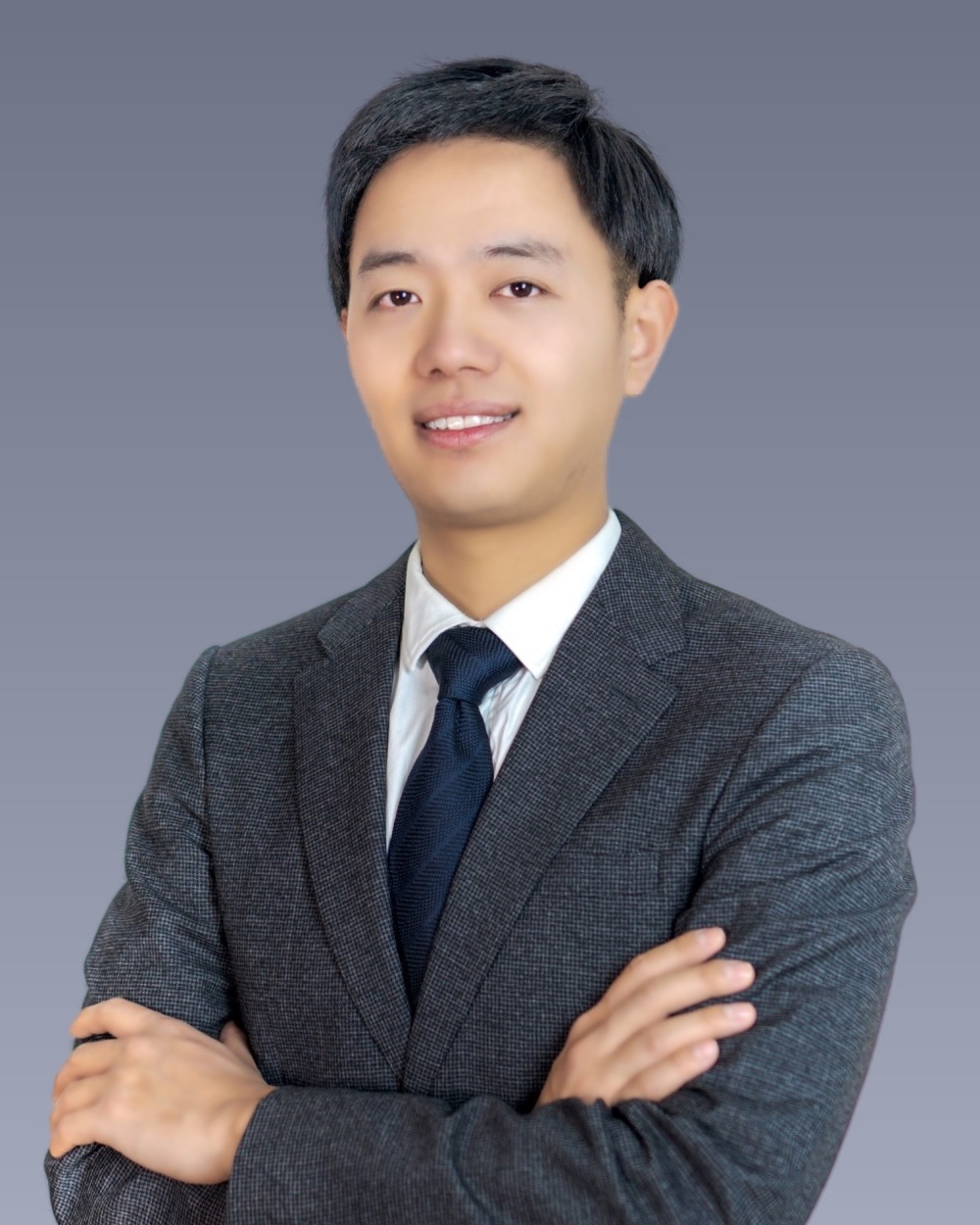}}]{Yihua Xu}
 received in 2012 his B.Sc. degree from Fuzhou University, M.Sc. in 2016 from University of Exeter, and in 2021 the Ph.D. degree from TU Kaiserslautern. Since June 2021 he starts to work as a quantitative analyst at KPMG Germany, and focuses on financial derivatives pricing and risk management. 
\end{IEEEbiography}

\begin{IEEEbiography}[{\includegraphics[width=1in,height=1.25in,clip,keepaspectratio]{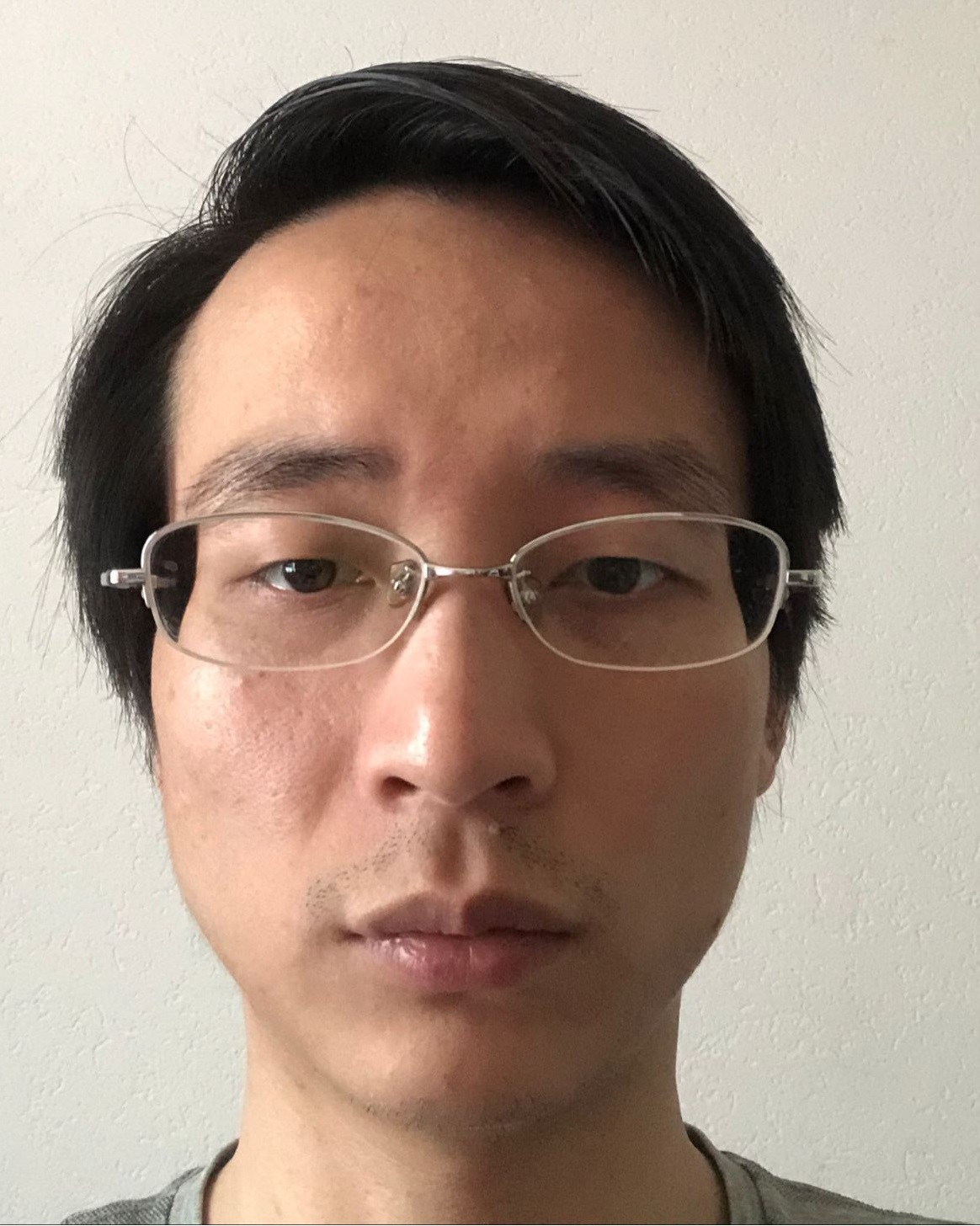}}]{Di Feng} received the M.A. in Economics from Yokohama National University and M.Res. in Economics from Universitat Aut\`onoma de Barcelona in 2016 and 2018, respectively. Currently, he is working for his Ph.D. degree in Economics at HEC - University of Lausanne. His research field covers decision theory, game theory and operations research, with a particular interest in market/mechanism design.
\end{IEEEbiography}

\begin{IEEEbiography}[{\includegraphics[width=1in,height=1.25in,clip,keepaspectratio]{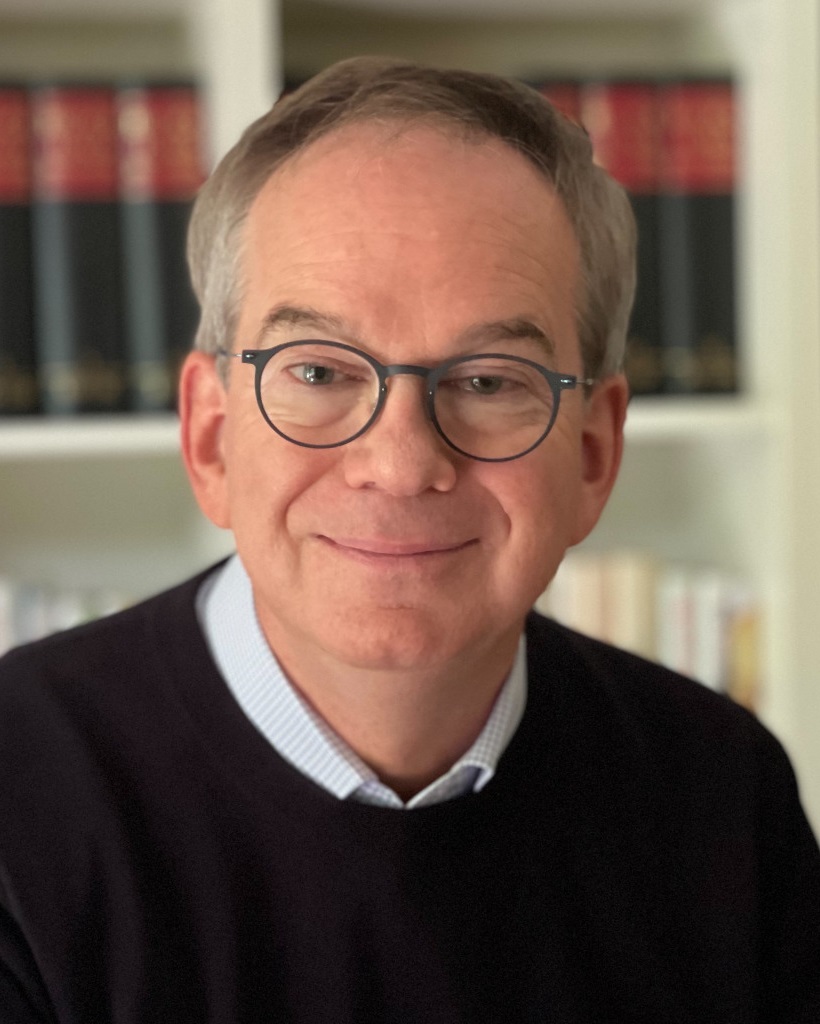}}]{Hans D. Schotten} (S'93--M'97) received the Ph.D. degree from the RWTH Aachen University of Technology, Germany, in 1997. From 1999 to 2003, he worked with Ericsson. From 2003 to 2007, he worked with Qualcomm. He became the Manager of a R\&D Group, a Research Coordinator for Qualcomm Europe, and the Director for Technical Standards. In 2007, he accepted the offer to become the Full Professor with the University of Kaiserslautern. In 2012, he became a Scientific Director of the German Research Center for Artificial Intelligence (DFKI) and the Head of the Department for Intelligent Networks. He served as the Dean of the Department of Electrical Engineering, University of Kaiserslautern from 2013 until 2017. He has authored more than 200 papers and participated over 40 European and national collaborative research projects. Since 2018, he has been the Chairman of the German Society for Information Technology and a Member of the Supervisory Board of the VDE.
\end{IEEEbiography}



\end{document}